\documentclass[aps, longbibliography,groupedaddress]{revtex4-1}
\pdfoutput=1
\usepackage{graphicx}
\usepackage{dcolumn}
\usepackage{bm}
\usepackage{verbatim}
\usepackage{amssymb}
\usepackage{amsfonts}
\usepackage{amsmath}
\usepackage{graphicx}
\usepackage{braket}
\usepackage{subcaption}

\DeclareMathOperator{\Ima }{Im}
\DeclareMathOperator{\Rea }{Re}

\begin{document}

\title{Evanescent-wave Johnson noise in small devices}
\author{Vickram N. Premakumar, Maxim G. Vavilov, and Robert Joynt}
\date{\today}

\begin{abstract}
In many quantum computer architectures, the qubits are in close proximity to
metallic device elements. The fluctuating currents in the metal give rise to
noisy electromagnetic fields that leak out into the surrounding region.
These fields are known as evanescent-wave Johnson noise. The noise can
decohere the qubits. We present the general theory of this effect for charge
qubits subject to electric noise and for spin and magnetic qubits subject to
magnetic noise. A mapping of the quantum-mechanical problem onto a problem
in classical electrodynamics simplifies the calculations. The focus is on
relatively simple geometries in which analytical calculations can be done.
New results are presented for the local noise spectral density in the
vicinity of cylindrical conductors such as small antennae, noise from
objects that can be treated as dipoles, and noise correlation functions for
several geometries. We summarize the current state of the comparison of
theory with experimental results on decoherence times of qubits. \ Emphasis
is placed on qualitative understanding of the basic concepts and phenomena.
\end{abstract}

\maketitle

\address{Department of Physics, University of Wisconsin, Madison, Wisconsin
53706, USA}

\section{Introduction}

The prospect of quantum computing has inspired many designs for the
manipulation of small coherent quantum systems - qubits. \ Qubits are often
located very near electrodes that contain many mobile charges and spins. \
The thermal and quantum motion of these charges and spins creates random
electromagnetic fields that can decohere the qubits, an effect strenuously
to be avoided. \ This noise is a species of Johnson noise. \ 

J.B. Johnson discovered this noise in 1927 in the course of a research
program to improve the performance of amplifiers \cite{johnson}. \ H.\
Nyquist soon explained it theoretically using ingenious applications of
equilibrium thermodynamics to thought experiments \cite{Nyquist}. \ When the
general relation of fluctuation and dissipation was discovered by H.B.
Callen and T.A. Welton in 1951, they regarded their fluctuation-dissipation
theorem (FDT) as a ``Generalized Nyquist Relation" \cite{Callen}. \ The
later, more general, theory of linear response of Kubo developed out of the
FDT \cite{Kubo}. \ \ This is an interesting example of important and general
basic science coming from research on very specific technological issues.

The Nyquist formula is 
\begin{equation}
\left\langle V^{2}\right\rangle _{\omega}=2k_{B}T~R\left( \omega\right)
\label{eq:nyquist}
\end{equation}
where 
\begin{equation*}
\left\langle V^{2}\right\rangle _{\omega}=\int_{-\infty}^{+\infty
}dt~e^{i\omega t}\left\langle V\left( 0\right) V\left( t\right)
\right\rangle .
\end{equation*}
Here $V$ is the voltage drop between the ends of a resistor with a possibly
frequency-dependent resistance $R.$ \ The angle brackets are an average over
the stationary random process that $V$ represents. \ The rms voltage noise $ 
\sqrt{\left\langle V^{2}\right\rangle _{\omega}}$ is the quantity usually
quoted (in units of volts per root Hertz), since it is often practical to
measure the drop with a bandpass filter in a frequency range where $R$ is
more or less constant. \ Johnson himself verified that this formula holds
independent of the shape, size, or constitution of the resistor. \ These
days, Eq. (\ref{eq:nyquist}) is recognized as the high-temperature limit of
the more general formula 
\begin{equation}
\left\langle V^{2}\right\rangle _{\omega}=\hbar\omega\coth\left( \hbar
\omega/2k_{B}T\right) ~R\left( \omega\right)  \label{eq:quantum}
\end{equation}
that follows from the quantum-mechanical version of the FDT. \ 
For applications to qubits we need a generalization of the Nyquist form of the FDT, which gives the voltage drop between two points in a resistor. In particular, we need a theory that works between any two points irrespective of whether they are on a
resistor; we would also like to understand the connection between the Nyquist relation with that other famous kind of thermal electromagnetic field - blackbody radiation. Quantum field theory gives the needed generalization. \ The main difficulty is to formulate finite-temperature quantum electrodynamics in such a way that the only inputs required are the macroscopic electric and magnetic response functions 
$\varepsilon \left( \vec{r},\omega \right) $ and $\mu \left( \vec{r},\omega
\right) .$ The outputs of the theory are the noise spectral densities, which are the field fluctuations at a single spatial point (sufficient to calculate the decoherence of point qubits), and the noise correlation functions which give the fluctuations at spatially separated points (required to calculate the decoherence of extended qubits). We will give precise definitions of these quantities below. The formalism required to do this was constructed in the 1950s by Lifshitz \cite{lifshitz} and
Rytov \cite{rytov} and the theory was further developed by Agarwal \cite 
{agarwal}. \ These authors built on earlier work of Casimir \cite{Casimir}.
\ An accessible treatment is given by Lifshitz and Pitayevskii \cite{LLSP2}.
\ There is a fairly large literature on the application of this formalism to
heat transfer and friction in small devices which has been reviewed by
Volokitin and Persson \cite{volokitin}.\ 

\begin{figure}[tbp]
\centering
\includegraphics[width=0.7\linewidth]{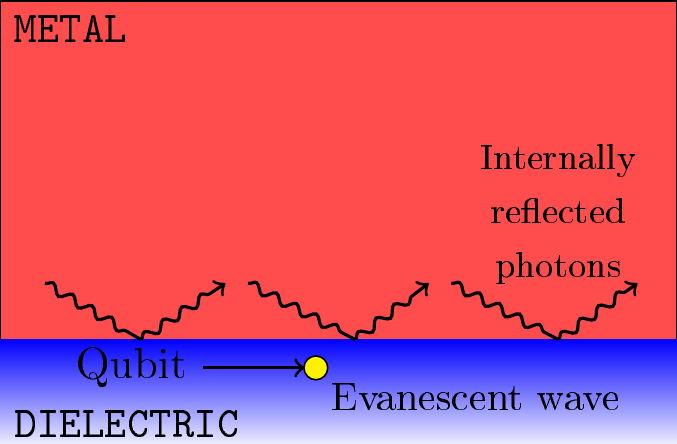}
\caption{}
\label{fig:evanescent}
\end{figure}

Before proceeding with the development of the formalism, we first give a
qualitative picture of how we expect noise to leak out of metallic device
elements, taking the lead from a paper of Pendry \cite{pendry}. \ Consider a
piece of metal surrounded by an insulator. \ For the sake of argument, let
us specify that the metal is hotter than its environment. \ The
Stefan-Boltzmann formula tells us that the total EM power radiated depends
only on the surface area and the temperature of the object, not on its
conductivity. \ The radiation is the result of photons thermally generated
in the metal leaking out through the surface. \ The metal has a dielectric
function $\varepsilon\left( \omega\right) =1+4\pi i\sigma/\omega,$ where the
conductivity $\sigma$ nearly always satisfies $\sigma/\omega\gg 1$ (and this
is true for all frequencies considered in this paper). \ $\left\vert
\varepsilon\right\vert $ is much greater than unity, so the speed of light
(to the extent that it can be defined for the highly overdamped modes of the
metal) is small relative to the surrounding insulator. \ This immediately
implies that the photon density of states and the equilibrium density
depends on $\sigma$. \ This presents a paradox, since the radiated power is
independent of $\sigma.$ \ This paradox is resolved by the realization that
a high photon density of states is always accompanied by a high probability
of internal reflection of the photon \cite{Neumann,Kha}. \ The cancellation
of these effects gives the universal coefficient of blackbody radiation.
However, internal reflection is always accompanied by an evanescent wave
(Fig. \ref{fig:evanescent}). \ This in turn implies that there will be
strong Johnson noise near a metallic surface for any material having $ 
|\varepsilon| \gg  1$. \ This is called evanescent-wave Johnson noise (EWJN).
\ This physical picture tells us that the proper treatment of boundary
conditions will be very important. \ This in turn implies that for ordinary,
non-magnetically active metals, the behavior of electric noise is quite
different from magnetic noise, since magnetic fields can penetrate those
materials much more easily. 

It is very important to distinguish between EWJN and the more commonly
discussed circuit Johnson noise (CJN). \ If we consider two separate
metallic elements in a small device, usually the path of least resistance
between them runs through the external circuit. \ \ Thus CJN is a physical
effect that involves two or more device elements that convey information
about the external circuit to the qubit. \ EWJN, in contrast, is an effect
that occurs even without the external circuit, and fundamentally arises from
individual device elements. \ CJN and EWJN thus come from different physical
sources. \ For the most part, they can be calculated separately and they are
basically additive. \ 

\ The implications of Johnson noise for decoherence of atomic qubits were
first discussed by Henkel and collaborators \cite{Henkel1999, HenkelWilkens} 
, in the context of heating of trapped ions by the walls of the trap. \ The
local noise spectral densities for both electric and magnetic fields
relevant to the situation of point qubits near a conducting half-space were
calculated and loss and decoherence rates were extracted. \ These
predictions were quantitatively verified in experiments that measured losses
from magneto-optical traps \cite{Harber2003}. \ The lifetimes in the
experiments are of order 10 s and the distances from the walls 10 to 100 $ 
\mu $m. \ At about the same time, other qubit applications were discussed by
Sidles et al. \cite{Sidles}. \ In semiconductor and some other solid-state
implementations of quantum computing, the distance scales are much less than
in the atom experiments and this suggests that the effects of Johnson noise
could be appreciable for those systems \cite{averin,Makhlin,Dykman,Huang, Poudel, Langsjoen2012}.\ Indeed, a recent experiment with a diamond
film containing NV centers on a silver substrate demonstrated decoherence of
qubits due to EWJN in a very direct and quantitative fashion \cite{Kolkowitz}. 

Charge quantum dot qubits displayed lifetimes in the range of $T_{1}\sim $
10 ns, which was shorter than expected based on decoherence mechanisms such
as coupling to phonons \cite{Hayashi,Fujisawa, Petta, Gorman}. 
This spurred theoretical work on CJN for double quantum dots \cite 
{Wilhelm2010}, and even though it appears that it cannot be the main
mechanism in this instance, the effects are still appreciable. \ 

There has been a small amount of work on the very interesting topic of noise
from micromagnets implanted in semiconductors \cite{Neumann,Kha}. \ However,
in this paper we shall deal only with non-magnetic materials, so the
magnetic permeability $\mu=1$ everywhere. \ \ 
In this paper, we focus exclusively on EWJN. We cover only
analytic calculations and physical considerations. \ Numerical calculations
on realistic devices are not included. \ To our knowledge, no such
calculations exist at present, though the calculations in Ref. \cite{Neumann} represent a start in this direction. \ 

The literature at present only contains analytic results for the half-space, single film \cite{Langsjoen2014}, and two-film geometries. In the next section we outline the basic formalism of EWJN. Sec. 2 describes how to apply the results to compute lifetimes of qubits.\ Sec. 3 gives the applications to electric field noise and decoherence of charge qubits. \ Sec. 4 is a parallel discussion for magnetic field noise and spin qubits. \ Sec. 5 gives the current situation with regard to comparison of theory and experiment. Sec. 6 gives a summary and describes the implications for future
qubit designs.

\ The overall structure of the paper is meant to reflect the logical development of the subject, with reasonably complete derivations of the main results.  If the reader's main concern is just with new results, then these are to be found as follows.
\begin{itemize}
	\item We present exact analogies to equivalent problems in classical electromagnetic theory that greatly simplify the calculations in Sec. $2.2 $. 
	\item We provide explicit results for
	the noise spectral density that determines the decoherence of point qubits for new geometries. We find that for the conducting cylinder and electrode, the spectral densities exhibit anisotropies. These anisotropies can serve as a sharp criterion for the presence of EWJN. They can also be exploited to substantially increase $T_1$ and $T_2$ by suitable qubit orientation. These results are found in Secs. $4.1-4.3$ and $5.1-5.3$.
	\item We compute noise correlation functions that are needed for the determination of decoherence times for extended qubits. These results are found in Secs. $4.1 - 4.3$ and $5.1 - 5.3$.
	\item We describe in detail how to apply these noise calculations to compute relaxation and decoherence times for qubits in the noise field. These results are found in Sec. $3.1$ and $3.2 $. 
\end{itemize} 
Overall, the comparisons with present experimental results indicate that EWJN is not the dominant relaxation mechanism for many charge qubit implementations. On the other hand, the observed relaxation time for certain spin qubits can be explained by the calculations we present here.

\section{General Formalism}

\subsection{Photon Green's Functions}

We consider a system at temperature $T$ and regard the dielectric function $ 
\varepsilon\left( \vec{r},\omega\right) $ as given. $\ $As stated above, $ 
\mu=1$ everywhere so $\vec{B}=\vec{H}.$ \ We will work in the realm of
macroscopic electrodynamics, i.e., all quantities are averaged over
distances of order $a,$ where $a$ is an interatomic distance. \ This
excludes a large class of physical situations that can be important in qubit
devices, namely those in which the noise sources are few in number or
otherwise cannot cannot be be considered as members of a continuum. The
results in this paper do not apply to such situations.\ 

Our derivation in this section follows Ref. \cite{LLSP2}. \ We present it
here to introduce the concepts and to establish notation. \ 

We shall work in the temporal gauge where the scalar potential $\phi=0.$ \
The retarded photon Green's function is 
\begin{equation*}
iG_{ij}\left( \vec{r},t;\vec{r}^{\prime},t^{\prime}\right) =\Theta\left(
t-t^{\prime}\right) \left\langle \widehat{A}_{i}\left( \vec{r},t\right) 
\widehat{A}_{j}\left( \vec{r}^{\prime},t^{\prime}\right) -\widehat{A}
_{j}\left( \vec{r}^{\prime},t^{\prime}\right) \widehat{A}_{i}\left( \vec {r} 
,t\right) \right\rangle .
\end{equation*}
Here $\Theta\left( x\right) =1$ if $x>0$ and $\Theta\left( x\right) =0$ if $ 
x<0.$ \ $i,j$ run over $x,y,z.$ \ The angle brackets represent a thermal
ensemble average. \ The $\widehat{A}$ are photon operators for the vector
potential in the interaction picture. \ Since for a closed system we have 
\begin{equation}
G_{ij}\left( \vec{r},t;\vec{r}^{\prime},t^{\prime}\right) =G_{ij}\left( \vec{ 
r},\vec{r}^{\prime},t-t^{\prime}\right)
\end{equation}
we define 
\begin{equation}
G_{ij}\left( \vec{r},\vec{r}^{\prime},\omega\right) =\int_{-\infty}^{\infty
}dt~e^{i\omega t}G_{ij}\left( \vec{r},t;\vec{r}^{\prime},0\right)
\end{equation}
and this function satisfies an Onsager relation: 
\begin{equation}
G_{ji}\left( \vec{r}^{\prime},\vec{r},\omega\right) =G_{ij}\left( \vec {r}, 
\vec{r}^{\prime},\omega\right) .
\end{equation}
Fortunately, we will not need to consider the operator properties of $ 
\widehat{A}$ in detail. \ Instead, we will derive a differential equation
for $G$. \ In the presence of a classical current $\vec{J},$ $H^{\prime}$ is
the perturbation to the free-space Maxwell Hamiltonian with 
\begin{equation}
H^{\prime}=-\frac{1}{c}\int J_{i}\left( \vec{r},t\right) ~A_{i}~d^{3}r,
\end{equation}
with a summation convention over Cartesian indices. \ The expectation value
of $\widehat{A}_{i}$ is given by the Kubo formula 
\begin{equation}
\left\langle \widehat{A}_{i}\left( \vec{r},\omega\right) \right\rangle =- 
\frac{1}{\hslash c}\int G_{ij}\left( \vec{r},\vec{r}^{\prime} ,\omega\right)
~J_{j}\left( \vec{r}^{\prime},\omega\right) ~d^{3}r^{\prime }.  \label{eq:ag}
\end{equation}
In the following, angle brackets and the argument $\omega$ will often be
omitted. \ \ \ 

In many situations electronic length scales such as the mean free path $\ell$
is much smaller than all the other lengths in the problem and consequently
there is a local relation between the electric displacement and the electric
field: $\vec{D}\left( \vec{r}\right) =\varepsilon\left( \vec{r}\right) $ $ 
\vec{E}\left( \vec{r}\right) $. Then Maxwell's equation is 
\begin{equation}
\nabla\times\vec{B}=\frac{4\pi}{c}\vec{J}-i\frac{\omega}{c}\varepsilon\left( 
\vec{r}\right) \vec{E},
\end{equation}
and the fields are given in this gauge by 
\begin{equation}
\vec{B}=\nabla\times\vec{A}~\text{and }\vec{E}=i\frac{\omega}{c}\vec{A}.
\end{equation}
Thus we have that 
\begin{equation}
-\nabla^{2}\vec{A}+\nabla\left( \nabla\cdot\vec{A}\right) -\frac{\omega
^{2}\varepsilon\left( \vec{r}\right) }{c^{2}}\vec{A}=\frac{4\pi}{c}\vec{J}
\end{equation}
which in index notation with the summation convention is 
\begin{equation}
\left[ -\delta_{ij}\left( \nabla^{2}+\frac{\omega^{2}\varepsilon\left( \vec{r 
}\right) }{c^{2}}\right) +\partial_{i}\partial_{j}\right] A_{i} =\frac{4\pi}{ 
c}J_{j}.  \label{eq:a}
\end{equation}
Applying the operator in square brackets to both sides of \ref{eq:ag} gives 
\begin{equation}
\frac{4\pi}{c}J_{j}\left( \vec{r}\right) =-\frac{1}{\hslash c}\int\left\{  
\left[ -\delta_{ij}\left( \nabla^{2}+\frac{\omega^{2}\varepsilon\left( \vec{r 
}\right) }{c^{2}}\right) +\partial_{i}\partial_{j}\right] G_{ik}\left( \vec{r 
},\vec{r}^{\prime}\right) \right\} ~J_{k}\left( \vec {r}^{\prime}\right)
~d^{3}r^{\prime}.
\end{equation}
Since this is true for any $J,$ it implies that 
\begin{equation}
\left[ -\delta_{ij}\left( \nabla^{2}+\frac{\omega^{2}\varepsilon\left( \vec{r 
}\right) }{c^{2}}\right) +\partial_{i}\partial_{j}\right] G_{ik}\left( \vec{r 
},\vec{r}^{\prime}\right) =-4\pi\hslash~\delta^{3}\left( \vec{r}-\vec{r} 
^{\prime}\right) \delta_{jk}.  \label{eq:g}
\end{equation}
The differential operators act on $\vec{r},$ not $\vec{r}^{\prime}.$ \ For a
fixed $\vec{r}^{\prime}$ (source position), this is an inhomogeneous partial
differential equation when $j=k$ and a homogeneous partial differential
equation when $j\neq k$ for the functions $G_{jk}$. \ Tangential $\vec{E}$,
normal $D=\varepsilon\vec{E}$ and $\vec{B}=\vec{H}$ are continuous at the
boundary between different media. \ We have that\ $E_{i}\left( \vec 	{r} 
\right) \sim\left( i\omega/c\right) G_{ij}\left( \vec{r},\vec 	{r} 
^{\prime}\right) $ and $B_{i}\left( \vec{r}\right) \sim\varepsilon
_{imn}\partial_{m}G_{nj}\left( \vec{r},\vec{r}^{\prime}\right) .$ Hence the
boundary conditions at a surface with a discontinuity in $\varepsilon\left( 
\vec{r}\right) $ with normal vector $\widehat{n}$ are: 
\begin{align*}
\epsilon_{ijk}n_{i}G_{jm}& ~~~ \text{ continuous for all }k,m \\
\varepsilon n_{i}G_{im}& ~~~ \text{ continuous for all }m  \\
\epsilon_{ijk}\partial_{i}G_{jm}& ~~~ \text{ continuous for all }k,m.
\end{align*}

Now assume that we can solve these differential equations and have the
response function $G.$ \ Then an application of the FD theorem yields 
\begin{align}
	\int &e^{i\omega\left( t-t^{\prime}\right) }\left\langle A_{i}\left( \vec {r} 
	,t\right) A_{j}\left( \vec{r}^{\prime},t^{\prime}\right) \right\rangle
	~d\left( t-t^{\prime}\right) \\
	&=\left\langle A_{i}\left( \vec{r}\right) A_{j}\left( \vec{r} 
	^{\prime}\right) \right\rangle _{\omega} \\
	& =-\coth\left( \frac{\hslash\omega}{2k_{B}T}\right) \times \Ima  
	G_{ij}\left( \vec{r},\vec{r}^{\prime},\omega\right) .
\end{align}

As we will see below, the relaxation of a charge qubit with level separation 
$\omega$ in the neighborhood of $\vec{r}$ and $\vec{r}^{\prime}$ will be
determined by a correlation function of the type 
\begin{equation}
\left\langle E_{i}\left( \vec{r}\right) E_{j}\left( \vec{r}^{\prime }\right)
\right\rangle _{\omega}=-\frac{\omega^{2}}{c^{2}}\coth\left( \frac{ 
\hbar\omega}{2k_{B}T}\right) \Ima G_{ij}\left( \vec {r},\vec{r} 
^{\prime},\omega\right) .  \label{eq:efield}
\end{equation}

The relaxation of a spin qubit with level separation $\omega$ in the
neighborhood of $\vec{r}$ and $\vec{r}^{\prime}$ will be determined by a
correlation function of the type 
\begin{align}
\left\langle B_{i}\left( \vec{r}\right) B_{j}\left( \vec{r}^{\prime }\right)
\right\rangle _{\omega}&=-\coth\left( \frac{\hslash\omega}{2k_{B} T}\right) 
\notag \\
&\times\epsilon_{ikm}\epsilon_{jnp}\partial_{k}\partial_{n}^{\prime } 
\Ima G_{mp}\left( \vec{r},\vec{r}^{\prime},\omega\right) .
\end{align}
We shall also have occasion to refer to the mixed correlation function 
\begin{align}
\left\langle E_{i}\left( \vec{r}\right) B_{j}\left( \vec{r}^{\prime }\right)
\right\rangle _{\omega} & = -\frac{i\omega}{c} \coth\left( \frac{ 
\hslash\omega}{2k_{B} T}\right)  \notag \\
& \times \epsilon_{jkl} \partial_k^{\prime } \Ima G_{il}(\omega,r,r^{ 
\prime }).
\end{align}

\subsection{Physical Analogy}

Physical intuition for the meaning of $G_{ik}\left( \vec{r},\vec{r}^{\prime
},\omega\right) ,$ and a practical calculation method, may be obtained by
noting the similarity of Eqs. \ref{eq:a} and \ref{eq:g}. \ Place a
fictitious point electric dipole $\vec{p}$ at the point $\vec{r}^{\prime}.$
\ \ The current is 
\begin{equation*}
\vec{J}^{\left( f\right) }\left( \vec{r}\right) =-i\omega\vec{p} 
~\delta^{3}\left( \vec{r}-\vec{r}^{\prime}\right) .
\end{equation*}
\ and the resulting fictitious vector potential and the electric field are
given by 
\begin{equation*}
\left[ \partial_{i}\partial_{l}-\delta_{il}\nabla^{2}-\delta_{il}~\frac{ 
\omega^{2}\varepsilon\left( \vec{r}\right) }{c^{2}}\right] ~A_{l}^{\left(
f\right) }\left( \vec{r}\right) =-i\omega p_{i}\frac{4\pi }{c} 
\delta^{3}\left( \vec{r}-\vec{r}^{\prime}\right)
\end{equation*}
and 
\begin{equation}
\left[ \partial_{i}\partial_{l}-\delta_{il}\nabla^{2}-\delta_{il}~\frac{ 
\omega^{2}\varepsilon\left( \vec{r}\right) }{c^{2}}\right] ~E_{l}^{\left(
f\right) }\left( \vec{r}\right) =p_{i}\frac{4\pi\omega^{2}}{c^{2}} 
\delta^{3}\left( \vec{r}-\vec{r}^{\prime}\right) .  \label{eq:e}
\end{equation}
On the other hand, multiplying Eq. (\ref{eq:g})
 by $p_{k}$ and summing over $k$
we find:

\bigskip 
\begin{equation}
\left[ \partial_{i}\partial_{l}-\delta_{il}\nabla^{2}-\delta_{il}~\frac{ 
\omega^{2}\varepsilon\left( \vec{r}\right) }{c^{2}}\right] ~G_{lk}\left(
\omega;\vec{r},\vec{r}^{\prime}\right) p_{k}=-4\pi
\hslash~p_{i}~\delta^{3}\left( \vec{r}-\vec{r}^{\prime}\right) .
\label{eq:d}
\end{equation}
Comparison of Eqs. (\ref{eq:e}) and (\ref{eq:d}) says that 
\begin{equation}
G_{lk}\left( \omega;\vec{r},\vec{r}^{\prime}\right) p_{k}=-\frac{\hslash
c^{2}}{\omega^{2}}E_{l}^{\left( f\right) }.  \label{eq:de}
\end{equation}
and, using Eq. (\ref{eq:efield}) 
\begin{equation}
\left\langle E_{i}\left( \vec{r}\right) E_{j}\left( \vec{r}^{\prime }\right)
\right\rangle _{\omega}~p_{j}=\hslash\coth\left( \frac {\hslash\omega}{ 
2k_{B}T}\right) \Ima E_{i}^{\left( f\right) }~\text{(no sum).}
\label{eq:enoise}
\end{equation}
Hence if we wish to find (say) $G_{xy},$ we solve the fictitious classical
problem of an oscillating dipole $\vec{p}=\left( 0,p_{y},0\right) $ at the
point $\vec{r}^{\prime}$ and compute $E_{x}^{\left( f\right) }$\bigskip\ at
the point $\vec{r}.$ \ Then 
\begin{equation}
G_{xy}\left( \vec{r},\vec{r}^{\prime}\right) =-\frac{\hslash c^{2}}{ 
\omega^{2}}E_{x}^{\left( f\right) }/p_{y}.  \label{eq:fic}
\end{equation}
\ We can compute all 9 components of $G$ in this way. There is a similar
analogy for magnetic fluctuations. \ The current of a point magnetic dipole $ 
\vec{m}$ at $\vec{r}^{\prime}$ may be written as 
\begin{equation*}
J_{i}^{\left( f\right) }\left( \vec{r}\right) =c~\varepsilon
_{ijk}~\partial_{j}\delta^{3}\left( \vec{r}-\vec{r}^{\prime}\right) ~m_{k}.
\end{equation*}
The equation for the vector potential in this situation is 
\begin{align}
& \left[ \partial_{i}\partial_{l}-\delta_{il}\nabla^{2}-\delta_{il}~\frac{ 
\omega^{2}\varepsilon\left( \vec{r}\right) }{c^{2}}\right] ~A_{l}^{\left(
f\right) }\left( \vec{r}\right) \\
&=-4\pi~\varepsilon _{ijk}~\partial_{j}^{\prime}\delta^{3}\left( \vec{r}- 
\vec{r}^{\prime}\right) ~m_{k}.  \label{eq:al}
\end{align}
Multiplying Eq. (\ref{eq:g}) by$~\epsilon_{kmn}~\partial_{m}^{\prime}m_{n}/ 
\hbar$ and summing over $m$ and $n$ we find 
\begin{align}
& \left[ -\delta_{ij}\left( \nabla^{2}+\frac{\omega^{2}\varepsilon\left( 
\vec{r}\right) }{c^{2}}\right) +\partial_{i}\partial_{j}\right] \frac {1}{ 
\hbar}\epsilon_{kmn}~m_{n}\partial_{m}^{\prime}G_{ik}\left( \vec {r},\vec{ 
r}^{\prime}\right) \notag \\
&=-4\pi\epsilon_{jmn}~\partial_{m}^{\prime }m_{n}\delta^{3}\left( \vec{r}- 
\vec{r}^{\prime}\right) .  \label{eq:am}
\end{align}

\bigskip Equating the curl of Eqs. (\ref{eq:al}) and (\ref{eq:am}) yields 
\begin{equation}
\frac{1}{\hbar}\epsilon_{ijk}\epsilon_{lmn}~m_{n}\partial_{i} 
\partial_{m}^{\prime}G_{jl}\left( \vec{r},\vec{r}^{\prime}\right)
=B_{k}^{\left( f\right) }\left( \vec{r},\vec{r}^{\prime}\right) .
\label{eq:db}
\end{equation}
Hence if we wish to find the magnetic correlations, we first solve the
fictitious classical problem of the magnetic field $\vec{B}^{\left( f\right)
}(\vec{r},\vec{r}^{\prime})$ at the point $\vec{r}$ resulting from an
oscillating point magnetic dipole $\vec{m}$ at the point $\vec{r}^{\prime}.$
\ For example, to find the physical magnetic field noise spectral density\
we place a point magnetic dipole $\vec{m}$ in the $j$th direction at $\vec {r 
}^{\prime},$ compute $B_{i}^{\left( f\right) }\left( \vec{r},\vec {r} 
^{\prime}\right) ,$ and then 
\begin{equation}
\left\langle B_{i}\left( \vec{r}\right) B_{j}\left( \vec{r}^{\prime }\right)
\right\rangle = \frac{\hbar}{m_{j}}\coth(\hbar\omega/2k_{B}T)\Ima B_{i}^{\left(
f\right) }\left( \vec{r},\vec{r}^{\prime}\right) .
\label{eq:bgbb}
\end{equation}

The Maxwell equations relate $\vec{E}$ at even orders in $\omega$ with $\vec{ 
B}$ at odd orders and vice versa, so the theory has two uncoupled sectors. \
This is the reason that we need the two separate analogies represented by
Eqs. (\ref{eq:de}) and (\ref{eq:db}).

In the fictitious problem, the equations satisfied by the fields in the
vacuum are 
\begin{align*}
\nabla^{2}\vec{E}^{\left( f\right) } & =0 \\
\nabla^{2}\vec{B}^{\left( f\right) } & =0 \\
\nabla\cdot\vec{E}^{\left( f\right) } & =4\pi\rho/\varepsilon_{d} \\
\nabla\cdot\vec{B}^{\left( f\right) } & =4\pi\vec{J}/c
\end{align*}
and in the metal we have 
\begin{align*}
\nabla^{2}\vec{E}^{\left( f\right) }+2i\delta^{-2}\vec{E}^{\left( f\right) }
& =0 \\
\nabla^{2}\vec{B}^{\left( f\right) }+2i\delta^{-2}\vec{B}^{\left( f\right) }
& =0 \\
\nabla\cdot\vec{E}^{\left( f\right) } & =0 \\
\nabla\cdot\vec{B}^{\left( f\right) } & =4\pi\vec{J}/c,
\end{align*}
in the quasistatic case$.$ The boundary conditions are that the tangential
component of $\vec{E}^{\left( f\right) }$ and $\vec{B}^{\left( f\right) }$
are continuous at the interface of dielectric and metal, while the normal
component $E_{n}^{\left( f\right) }$of $\vec{E}^{\left( f\right) }$
satisfies $\left( 4\pi i\sigma/\omega\right) E_{n}^{\left( f\right)
}(m)=\varepsilon_{d}E_{n}^{\left( f\right) }\left( d\right) ,$ where $ 
E_{n}^{\left( f\right) }(m),E_{n}^{\left( f\right) }(d)$ is the normal
component of $E$ in the metal (respectively, the dielectric) as the surface
is approached. $\sigma$ is the DC conductivity of the metal. $ 
\varepsilon_{d} $ is the dielectric constant in the dielectric material.

\subsection{Quasistatic Approximation}

The subject of this paper is the random electric and magnetic fields that
decohere qubits in the neighborhood of small metallic objects. \ The
characteristic frequencies for the decoherence rarely exceed a few GHz, so
we restrict our attention to frequencies at or below this range. \ For this
reason we employ the quasistatic approximation from the start, setting the
vacuum wavevector $k=\omega/c=0.$ \ In the interior of a metal object with
conductivity $\sigma$ the characteristic length scale of the fields is the
skin depth $\delta=c/\sqrt{2\pi\sigma\omega}.$ \ The inverse skin depth $ 
\delta^{-1}=c/\sqrt{2\pi\sigma\omega}$ is proportional to $\sqrt{\left(
\sigma/\omega\right) }k\gg k$ and it is retained in the theory. \ For example,
the term $\omega^{2}\varepsilon\left( \vec{r}\right) /c^{2}$ in Eq. (\ref 
{eq:a}) can be neglected when $\vec{r}$ is in the dielectric or vacuum where $ 
\varepsilon\sim1$ but not when $\vec{r}$ is in the metal. \ In this
approximation, radiation fields are neglected.\ \ We assume that the Drude
model is a good approximation for the metals in question, and that $ 
\omega\ll  1/\tau,$ where $\tau$ is the relaxation time of electrons in the
metals. \ The dielectric function is always approximated as $ 
\varepsilon=4\pi i\sigma/\omega.$

\subsection{Nonlocal Effects}

This paper focuses on the cases where local response is valid. \ Roughly
speaking, this is when the distance of $\vec{r}$ and $\vec{r}^{\prime}$ from
the nearest metal surface is greater than the electron mean free path in the
metal. \ However, when the distance to the metal tends to zero, the local
expressions for noise strengths diverge, which is clearly unphysical. \ For
completeness, we briefly outline how to include nonlocality in the theory. \
Generally $\vec{D}\left( \vec{r}\right) ,$ the electric displacement,
depends on $\vec{E}\left( \vec{r}^{\prime}\right) $ according to $ 
D_{i}\left( \vec{r},t\right) =\int d^{3}r^{\prime}$ $\varepsilon _{ij}\left( 
\vec{r}-\vec{r}^{\prime},t-t^{\prime}\right) E_{j}\left( \vec{r} 
^{\prime},t^{\prime}\right) $ and when Fourier transformed this becomes $ 
D_{i}\left( \vec{k},\omega\right) =\varepsilon_{ij}\left( \vec {k} 
,\omega\right) E_{j}\left( \vec{k},\omega\right) $. $\ $Eq. (\ref{eq:g})
becomes 
\begin{align}
\left( -\delta_{ij}\nabla^{2}+\partial_{i}\partial_{j}\right)& G_{ik}\left( 
\vec{r},\vec{r}^{\prime}\right) -\delta_{ij}\frac{\omega^{2}}{c^{2}}\int
d^{3}r^{\prime\prime}\varepsilon_{im}\left( \vec{r},\vec{r}^{\prime\prime
}\right) G_{mk}\left( \vec{r}^{\prime\prime},\vec{r}^{\prime}\right) \\
&=-4\pi\hslash\delta^{3}\left( \vec{r}-\vec{r}^{\prime}\right) \delta_{jk}.
\end{align}
Use of this equation with an appropriate choice for $\varepsilon\left( \vec{r 
},\vec{r}^{\prime\prime}\right) $ cures the unphysical divergence at small
distances. In practice, to date only the problems of a conducting half-space
and conducting films have been treated using the nonlocal formalism \cite 
{Langsjoen2012,Poudel,Langsjoen2014}.

\section{Application to Qubits}

\subsection{Relaxation}

A qubit system in a noisy environment is described by a Hamiltonian $H = H_q
+ H_n(t)$ where $H_q$ admits two eigenstates $\ket{0}, \ket{1}$ such that $ 
H_q \ket{i} = \epsilon_i\ket{i}$. The relaxation rate for such a qubit in
the presence of EWJN is given by the Golden Rule-type formula 
\begin{equation}
\frac{1}{T_1} = \frac{1}{\hbar^2} \int_{-\infty}^{\infty} \overline{ 
	\braket{0|H_n(t)|1} \braket{1| H_n(0) |0}} e^{ -i \omega t} dt
\label{eqn:t1}
\end{equation}
Consider a qubit with charge, mass, and g-factor $e$, $m$, and $g$
respectively placed in a time dependent electromagnetic field described by $ 
\vec{A}(r,t)$. The full Hamiltonian is 
\begin{align*}
H &= \frac{1}{2m} \left( \vec{\Pi} - \frac{e}{c} \vec{A} \right)^2 + V( \vec{ 
	r}) - \frac{e g}{2m}\vec{B} \cdot \vec{S},
\end{align*}
where $\vec{\Pi} = - i \hbar \nabla$. Here we will restrict ourselves to $ 
\mathcal{O}(e)$ so the Hamiltonian can be written 
\begin{equation}
H = \frac{\vec{\Pi}^2}{2m} + V(\vec{r}) - \frac{e}{2mc} \left( \vec{\Pi}
\cdot \vec{A} + \vec{A} \cdot \vec{\Pi} \right) - \frac{e g}{2m} \vec{B}
\cdot \vec{S}.
\end{equation}
Imposing the gauge condition $\phi = 0$ we find a Hamiltonian readily
treated in the interaction picture. The charge distribution generating the
noise is contained in the metal, so at a nearby qubit we have $\nabla \cdot
E = \nabla^2 \phi + \frac{1}{c} \partial_t (\nabla \cdot \vec{A}) = 0$. For
finite frequency noise, this implies $\nabla \cdot \vec{A} = 0$, and thus $[ 
\vec{\Pi}, \vec{A}(\vec{r})] = - i \hbar \nabla \cdot \vec{A}(\vec{r}) = 0$.
The time dependence of the system is entirely due to the
electromagnetic noise and the static Hamiltonian $H_q = \frac{\vec{\Pi}^2}{2m 
} + V(\vec{r})$ . We are left with 
\begin{equation*}
H = H_q + H_n(t)
\end{equation*}
\begin{equation}
H_n(t) = - \frac{e}{mc} \vec{A}(r, t) \cdot \vec{\Pi} - \frac{e g}{2mc}\vec{B 
}(r,t) \cdot \vec{S}.  \label{eq:hn}
\end{equation}
Our interaction Hamiltonian can be written as a spatial Taylor series as
follows 
\begin{align*}
H_n(t) = &- \frac{e}{mc} \left[ A_i(0,t) + \left(\nabla_j
A_i(r,t)\right)_{r=0} r_j + \dots \right] \Pi_i \\
&- \frac{e g}{2mc} B_i S_i.
\end{align*}
This allows us to treat the relevant matrix elements term by term in
multipole moments, as described in \cite{Raab}. Truncating the series at
second order and evaluating the off-diagonal matrix elements gives us 

\begin{equation}
\frac{1}{T_1^{E}}= \frac{1}{\hbar^2} \braket{p_i} \braket{p_l}^* \braket{E_i E_l}_{\omega} \; \; \; ; \; \; \;   \frac{1}{T_1^{B}}= \frac{1}{\hbar^2} \braket{m_i} \braket{m_l}^* \braket{B_i B_l}_{\omega} \label{eq:T1exp}
\end{equation} 
\[  \frac{1}{T_1^{cross}}= \frac{1}{\hbar^2}\left( \braket{p_i} \braket{m_n}^* \braket{E_i B_n}_{\omega} + \braket{m_k} \braket{p_l}^* \braket{B_k E_l}_{\omega} \right).\]
	
Above we set $\hbar \omega = \epsilon_1
- \epsilon_0$ via Eq. (\ref{eqn:t1}). For brevity we also use $\braket{x} \equiv \braket{0|x|1}$ and $\braket{F_i(t) F_j(0)}_\omega = \braket{F_i F_j}_\omega.$ Here we only include dipole contributions; higher order multipole moments and more details of the calculation are treated in the appendix. 

In the case of the spin qubit the states $\ket{0}, \ket{1}$ are up and down
states of the spin part of the wavefunction. Hence $\ket{0} = \ket{\psi_0}
\otimes \ket{\uparrow}$ where $\ket{\psi_0}$ is the orbital part of the
wavefunction which is common to both states of the spin qubit. Immediately
we see that all the spatial operator matrix elements $\braket{p_i} =  
\braket{q_{ij}} = \braket{l_i} = 0$. Hence the above expression simplifies
to 
\begin{equation}
\frac{1}{T_1} = \frac{1}{\hbar^2} \left(\frac{eg}{2m}\right)^2 \braket{ S_k}  
\braket{S_n}^* \braket{B_k B_n}_\omega.
\end{equation}
For concreteness, suppose the spin qubit is localized in space and arranged
so the up and down states are eigenstates of $S_z$, then we can explicitly
compute the matrix elements to find 
\begin{align}
\frac{1}{T_1} &= \left(\frac{e g}{4m}\right)^2 \left( \braket{B_x B_x} 
_\omega + \braket{B_y B_y}_\omega \right).
\end{align}

\subsection{Dephasing}

Qubit relaxation is due to the off-diagonal matrix elements $\left\langle
0\right\vert H_{n}\left\vert 1\right\rangle $ and$~\left\langle 1\right\vert
H_{n}\left\vert 0\right\rangle $ of the noise Hamiltonian. \ The diagonal
elements $\left\langle 0\right\vert H_{n}\left\vert 0\right\rangle $ and$ 
~\left\langle 1\right\vert H_{n}\left\vert 1\right\rangle $ produce
dephasing. \ If the initial state is $\left( 1/\sqrt{2}\right) \left[
\left\vert 0\right\rangle +\left\vert 1\right\rangle \right] $, and the
state at time $t$ is $\left( 1/\sqrt{2}\right) \left[ \left\vert
0\right\rangle +e^{i\phi\left( t\right) }\left\vert 1\right\rangle \right] $
then $\phi$ is random after a time $T_{2}.$ \ The basic formulas for $T_{2}$
are as follows. \ We have

\begin{equation*}
\frac{1}{T_{2}}=\frac{1}{2T_{1}}+\frac{1}{T_{\phi}}.
\end{equation*}
where $T_{\phi}$ is the dephasing time. For a Johnson-type noise mechanism,
the Gaussian approximation for $T_{\phi}$ should be very accurate, since
many modes of the metal contribute to the noise. $\ T_{\phi}$ is then
calculated in the following way. Again let the applied field be in the $i$th
direction. The initial condition is $\phi\left( t=0\right) =1.$ \ We then
repeatedly measure $X=\left\vert 0\right\rangle \left\langle 0\right\vert
+\left\vert 1\right\rangle \left\langle 1\right\vert ,$ average to get $ 
\overline{X\left( t\right) }$ and the function $\Gamma_{i}\left( t\right) $
is defined by 
\begin{equation*}
\overline{X\left( t\right) }=\exp\left[ -\Gamma\left( t\right) \right]
X\left( 0\right) \cos\omega t.
\end{equation*}
and the Gaussian result for $\Gamma\left( t\right) $ is 
\begin{equation}
\Gamma\left( t\right) =\frac{t^{2}}{2}\int_{-\infty}^{\infty}~d\omega
~S\left( \omega\right) \frac{\sin^{2}\left( \omega t/2\right) }{\left(
	\omega t/2\right) ^{2}},  \label{eq:t2g}
\end{equation}
with 
\begin{align}
S\left( \omega\right)&=\frac{1}{\hbar^{2}}\int_{-\infty}^{\infty }dt~ 
\overline{\left[ \left\langle 1\right\vert H_{n}\left( t\right) \left\vert
	1\right\rangle -~\left\langle 0\right\vert H_{n}\left( t\right) \left\vert
	0\right\rangle \right]}  \notag \\
&\times \overline{\left[ \left\langle 1\right\vert H_{n}\left( 0\right)
	\left\vert 1\right\rangle -~\left\langle 0\right\vert H_{n}\left( 0\right)
	\left\vert 0\right\rangle \right]}~e^{-i\omega t}.  \label{eq:t2s}
\end{align}
Evidently we need the diagonal matrix elements of the time-dependent part of
the Hamiltonian from Eq. (\ref{eq:hn}).  Defining moments $p_i = e r_i$ and $m_i = 
\frac{e}{2mc} (l_i + g S_i)$. 
\begin{align*}
\braket{1|H_n(t)|1} - \braket{0|H_n(t)|0} = &- B_k(t) \left(\braket{m_k}_1 -  
\braket{m_k}_0 \right) \\
&+ E_k(t) \left(\braket{p_k}_1 - \braket{p_k}_0 \right)
\end{align*}
To keep things short let $\Delta x= \left(\braket{1|x|1} - \braket{0|x|0}
\right) $ for any operator $x$. The integral kernel becomes
\begin{align}
S(\omega) = \frac{1}{\hbar^2} &\left[\braket{B_iB_j}_\omega \Delta m_i
\Delta m_j -\braket{B_iE_j}_\omega \Delta m_i \Delta p_j \right.  \notag \\
-&\left.\braket{E_iB_j}_\omega \Delta p_i \Delta m_j + \braket{E_iE_j} 
_\omega \Delta p_i \Delta p_j \right].  \label{eq:t2exp}
\end{align}
In order to make use of Eq. (\ref{eq:t2g}) we need to make some mild
assumptions on the frequency dependence of the noise spectral density terms.
We can write 
\begin{equation*}
\Gamma\left( t\right) =t^{2}\int_{0}^{1/\tau}~d\omega~f\left( \omega\right)
~\omega\coth\left( \frac{\hbar\omega}{2k_{B}T}\right) \frac{\sin^{2}\left(
	\omega t/2\right) }{\left( \omega t/2\right) ^{2}}.
\end{equation*}
Again, $f\left( \omega\right) $ contains all the information about
conductivity, qubit position, device geometry, etc., but it depends weakly
on frequency at low frequency, and here we will take it to be independent of
frequency $f\left( \omega\right) =f_{0}$ until it falls rapidly to zero at $ 
\omega=1/\tau,$ where $\tau$ is the electron relaxation time. \ \ We note
first that at very short times ($t\ll \tau,\hbar/k_{B}T)$ we always get $ 
\Gamma\left( t\right) \sim t^{2}/t_{0}^{2}$ (Gaussian decay), where 
\[ t_0^2 = \frac{4 \tau^2}{f_0} \tanh \left( \frac{\hbar}{2 k_B T \tau}\right) .\]
As a result, Gaussian decay is only observed when the noise is quasi-static. Exponential decay at longer times
is the most important from the standpoint of EWJN. \ This is where $t\gg \tau$ and $t\gg $ $\hbar/k_{B}T$
and then we can write 
\begin{align*}
\Gamma\left( t\right) & =4f_{0}\int_{0}^{t/2\tau}~dx~\coth\left( \frac{\hbar
	x}{k_{B}Tt}\right) \frac{\sin^{2}x}{x} \\
& \approx 2\pi f_{0}~\frac{k_{B}Tt}{\hbar}.
\end{align*}
Hence, at any experimentally accessible temperature 
\begin{equation}
\frac{1}{T_{\phi}}=\frac{2\pi f_{0}k_{B}T}{\hbar}.
\label{eq:tph}
\end{equation}

\begin{figure}
	\centering
	\includegraphics[width=0.7\linewidth]{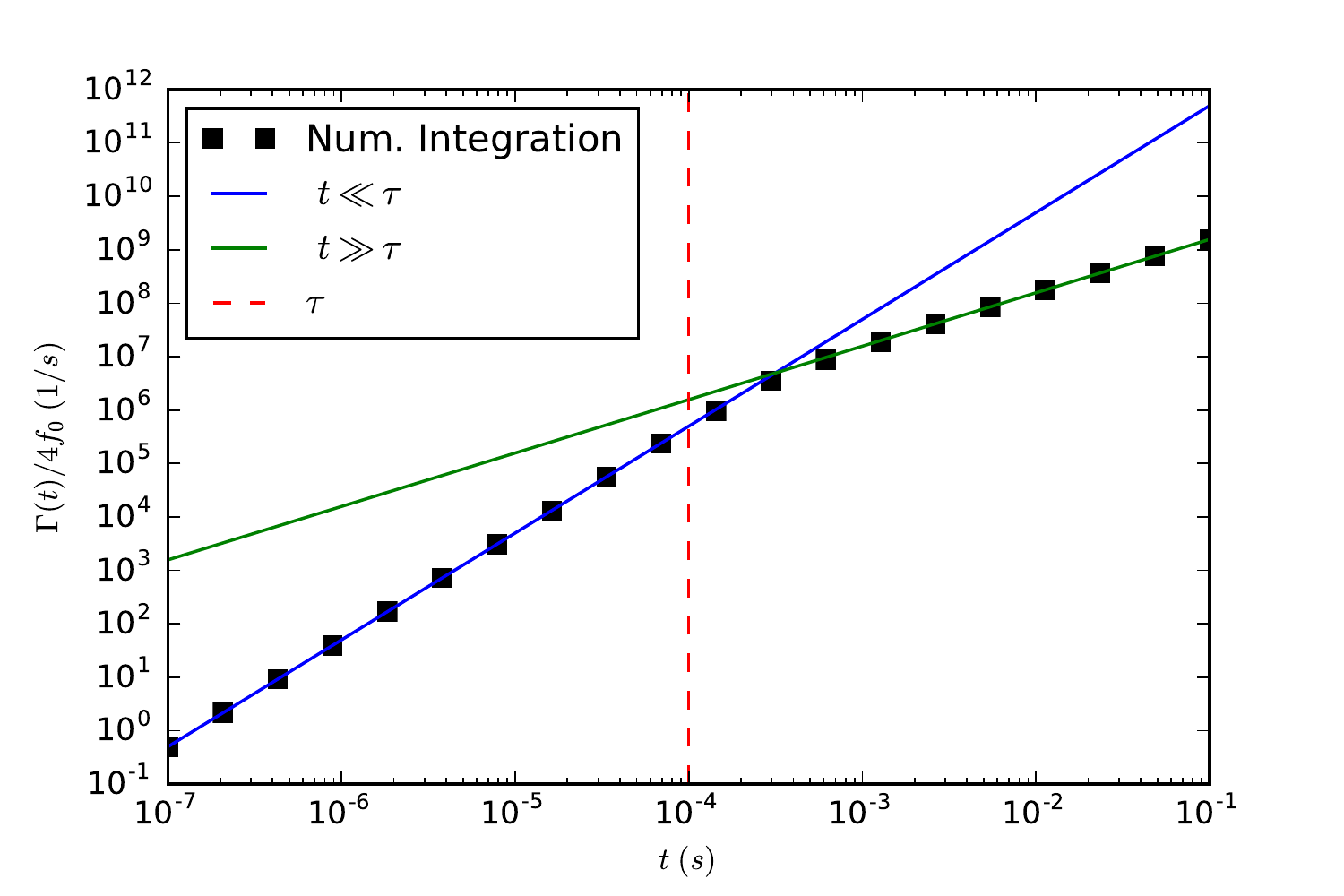}
	\caption{Eq. (\ref{eq:t2g}) for $S(\omega) = f_0 \; \omega \coth \left(\frac{\hbar \omega}{2 k_B T}\right)$ at $T = 0.1$ K and $\tau = 0.1 $ ms plotted alongside approximate results for Gaussian noise $(t \ll \tau)$ and exponential decay $(t\gg \tau )$.   }
	\label{fig:214dephasingratexover}
\end{figure}

We see that only off diagonal elements of the multipole moments determine $ 
T_1$ and all of the matrix elements come into the determination of $T_2$. If
the expectation values of the multipole moments are not significantly
different between the ground and excited qubit states ${T_\phi^{-1}}$ will
be small and ${T_2}^{-1} \approx (2 T_1)^{-1}$. Even if not, $T_1$ and $T_2$
will generally be of the same order of magnitude, which distinguishes EWJN
from many other noise mechanisms.

In many experiments, it appears that the noise spectrum has two components, a ``$1/f$" component that dominates at low frequencies, and a white component that is bigger at high frequencies. Using the qubit as a spectrometer \cite{Bylander} it has been shown that this happens both in GaAs devices \cite{Yacoby} and in Si devices \cite{Muhonen}. Echo techniques can mitigate the low-frequency noise but not the more pernicious white part. $T_2^{echo}$ , the decoherence time after echoing, can serve as a diagnostic for EWJN in this situation. The experiment of Ref. \cite{Yacoby} is particularly interesting in this regard, since it shows that the white component of the noise has a strong temperature dependence which the $1/f$ part is largely temperature ($T$) independent, strongly suggesting different origins for the two types of noise. However, $T_2^{echo}$ was proportional to $T^{-2}$ , while Eq. (\ref{eq:tph}) would predict a $T^{-1}$ behavior.

\begin{figure*}[tbp]
	\centering
	\begin{subfigure}[b]{.3\textwidth}
		\centering
		\includegraphics[height=1.2in]{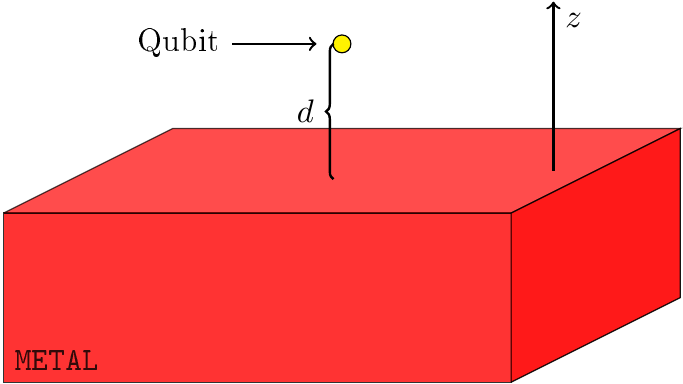}
		\caption{Half Space}
		\label{fig:halfsp}
	\end{subfigure}
	\begin{subfigure}[b]{.3\textwidth}
		\centering
		\includegraphics[height=2in]{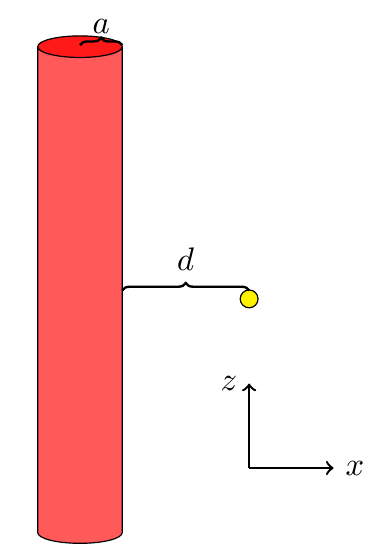}
		\caption{Cylinder}
		\label{fig:cyl}
	\end{subfigure}
	\begin{subfigure}[b]{.3\textwidth}
		\centering
		\includegraphics[height=2in]{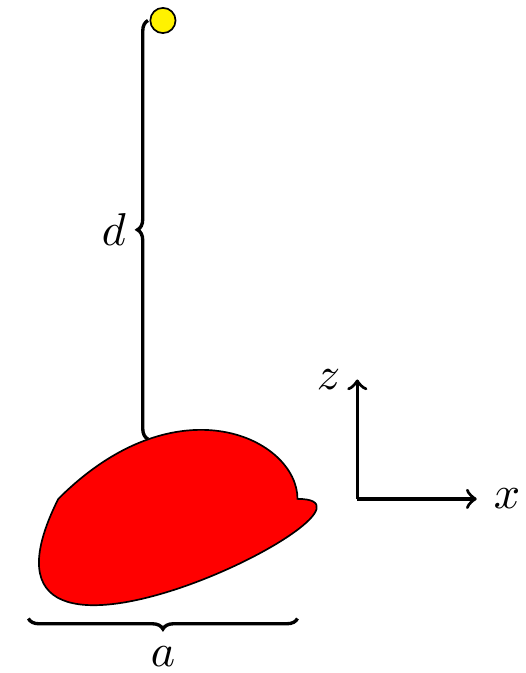}
		\caption{Distant Object}
		\label{fig:dist}
	\end{subfigure}
	\caption{Various qubit system geometries treated in this paper}
	\label{fig:geom}
\end{figure*}
\section{Electric noise}

The noise spectral density $\left\langle E_{i}\left( \vec{r}\right)
E_{j}\left( \vec{r}^{\prime}\right) \right\rangle _{\omega}$ generally
involves four length scales: $\left\vert \vec{r}-\vec{r}^{\prime}\right\vert
,$ the distance over which the correlations are to be measured; $d$, the
distance from the qubit to the conducting object(s); $\delta=c/\sqrt { 
2\pi\sigma\omega},$ the skin depth in the conductor(s); and $L,$ the linear
size of the conducting object(s). \ In most cases, the size of the qubit is
small, which means that usually the case $\vec{r}\approx\vec{r}^{\prime}$ is
of interest, and $\left\vert \vec{r}-\vec{r}^{\prime}\right\vert $ is the
smallest length in the problem. \ However, qubits can also be extended
objects, so we will give formulas as a function of $\vec{r}-\vec{r}^{\prime}$
where possible. \ As stated above, the vacuum wavelength is always taken to
be infinite. The simple geometries treated in this paper are shown in Fig.  
\ref{fig:geom}.

We will focus first on some limiting cases in which at least one of the
other three lengths is very different from the two others.
\subsection{Half Space}
\subsubsection{Point Qubit \ }

We first focus on some simple methods to compute $G_{ij}\left( \vec{r},\vec{r 
}^{\prime}=\vec{r}\right) =\delta_{ij}G_{ii}\left( \vec{r},\vec {r}^{\prime}= 
\vec{r}\right) ,$ which is sufficient for the calculation of the deocherence
of a point qubit. \ This case lends itself to some simple approximations
that are physically illuminating

\paragraph{Image Regime}

To understand this problem physically, we first outline the solution when $ 
d\ll \delta,$ since the problem is then essentially elementary. \ \ The
greater part of the electric field is concentrated within a sphere of radius
of order $d$ of the dipole. \ This implies that inside the metal we have
that $\nabla^{2}\vec{E}=\left( -2i/\delta^{2}\right) \vec{E}\approx0,$ since
the skin depth $\delta$ may be taken to be large. \ The problem now reduces
to the image problem for a static point charge in a medium with dielectric
constant $\varepsilon_{d}$ located at a distance $d$ from a half space with
dielectric constant $\varepsilon_{m}\approx4\pi i\sigma/\omega.$ \ For $z>0$
we have the equations $\nabla\cdot\vec{E}=4\pi\rho=4\pi\delta^{3}\left( \vec{ 
	r}-\vec {r}^{\prime}\right) $ and $\nabla\times\vec{E}=0.$ \ For $z<0$, we
have $\nabla\cdot\vec{E}=0$ and $\nabla\times\vec{E}=0.$ \ At the interface
we have \ $\varepsilon_{d}E_{z}\left( z=0_{+}\right)
=\varepsilon_{m}E_{z}\left( z=0_{-}\right) $ and $E_{x,y}\left(
z=0_{+}\right) =E_{x,y}\left( z=0_{-}\right) .$ \ This is the textbook\
image problem. \ Hence the solution for $z>0$ is given by $E=-\nabla\Phi,$
with $\Phi_{1}\left( \vec{r}\right) =q/\left\vert \vec{r}-\vec{r} 
^{\prime}\right\vert +q^{\prime}/\left\vert \vec{r}-\vec{r} 
^{\prime\prime}\right\vert $ and for $z<0$ by $\Phi_{2}\left( \vec{r}\right)
=q^{\prime\prime}/\left\vert \vec{r}-\vec{r}^{\prime }\right\vert .$ \ Here $ 
q^{\prime}=-q\left[ \left( \varepsilon _{m}-\varepsilon_{d}\right) /\left(
\varepsilon_{m}+\varepsilon_{d}\right) \right] $ and $q^{\prime\prime}=q 
\left[ \left( 2\varepsilon_{m}\right) /\left(
\varepsilon_{m}+\varepsilon_{d}\right) \right] .$ This satisfies the
differential equations and the boundary conditions. \ \ Hence the textbook
image solution carries over to this case.

We will do the $\left\langle E_{x}\left( \vec{r}\right) E_{x}\left( \vec {r} 
\right) \right\rangle _{\omega}$ correlation function first, so we place a
fictitious dipole $\vec{p}=p\widehat{x}$ at $\vec{r}^{\prime}=\left(
0,0,d\right) .$ \ Then we need the induced field at $\vec{r}.$ \ It is
produced by the image dipole $\vec{p}^{\prime}$ at $\vec{r}^{\prime\prime}:$

\begin{equation*}
p^{\prime}=-p\frac{\varepsilon_{m}-\varepsilon_{d}}{\varepsilon_{m}
	+\varepsilon_{d}}\approx-p\left( 1+\frac{i\omega\varepsilon_{d}}{2\pi\sigma } 
\right)
\end{equation*}
\ and the field from this charge is 
\begin{align*}
E_{x}^{\prime\left( f\right) }\left( \vec{r}\right) & =p^{\prime}\frac{ 
	3\left( \vec{r}-\vec{r}^{\prime\prime}\right) _{x}\left( \vec{r} -\vec{r} 
	^{\prime\prime}\right) _{x}-\left\vert \vec{r}-\vec{r}^{\prime\prime
	}\right\vert ^{2}}{\left\vert \vec{r}-\vec{r}^{\prime\prime}\right\vert ^{5}}
\\
& =p\left( 1+\frac{i\omega\varepsilon_{d}}{2\pi\sigma}\right) \frac {1}{ 
	\left( 2d\right) ^{3}}
\end{align*}
so 
\begin{equation*}
G_{xx}\left( \vec{r},\vec{r},\omega\right) =-\frac{\hbar c^{2}}{\omega ^{2} 
}\left( 1+\frac{i\omega\varepsilon_{d}}{2\pi\sigma}\right) \frac {1}{\left(
2d\right) ^{3}}
\end{equation*}
and using Eq. (\ref{eq:de}) we find at the position $\vec{r}=\vec{r} 
^{\prime}$ of the qubit that the physical local noise spectral density is 
\begin{align}
\left\langle E_{x}\left( \vec{r}\right) E_{x}\left( \vec{r}\right)
\right\rangle _{\omega}=\hbar\frac{\omega\varepsilon_{d}}{16\pi\sigma d^{3}}\coth\left( \frac{ 
	\hbar\omega}{2k_{B}T}\right) .  \label{eximage}
\end{align}
\ which at low temperatures $k_{B}T\ll \hbar\omega$ reduces to 
\begin{equation*}
\left\langle E_{x}\left( \vec{r}\right) E_{x}\left( \vec{r}\right)
\right\rangle _{\omega}=\hbar\frac{\omega\varepsilon_{d}}{16\pi\sigma d^{3} 
},
\end{equation*}
and at high temperatures $k_{B}T\gg \hbar\omega$ to 
\begin{equation*}
\left\langle E_{x}\left( \vec{r}\right) E_{x}\left( \vec{r}\right)
\right\rangle _{\omega}=\frac{k_{B}T\varepsilon_{d}}{8\pi\sigma d^{3}}.
\end{equation*}
Of course cylindrical symmetry implies that $\left\langle E_{y}\left( \vec {r 
}\right) E_{y}\left( \vec{r}\right) \right\rangle _{\omega}=\left\langle
E_{x}\left( \vec{r}\right) E_{x}\left( \vec{r}\right) \right\rangle
_{\omega}.$

It is important to note that the electric noise is \textit{inversely}
proportional to $\sigma.$ \ For really good metals, the screening is
complete and there is no dissipation and therefore no fluctuations in the
field. \ It is a general result that the result for $\vec{E}^{\left(
	f\right) }$ depends only on the \textit{ratio} of dielectric constants in
the two media, that is, on $\left( 4\pi i\sigma/\omega\right)
/\varepsilon_{d}.$ \ This follows immediately from inspection of the
boundary condition, which is the only place that $\varepsilon_{d}$ enters
the calculation. \ The $d^{-3}$ dependence follows immediately from the
physical analogy to the image problem.

Now we will do the $\left\langle E_{z}\left( \vec{r}\right) E_{z}\left( \vec{ 
	r}^{\prime}\right) \right\rangle _{\omega}$ correlation function, so we
place a dipole $\vec{p}=p\widehat{z}$ at $\vec{r}^{\prime}=\left(
0,0,d\right) .$ \ Then we need the induced field at $\vec{r}.$ \ The
calculation proceeds as for the x direction except for a change in sign of
the fictitious image dipole $\vec{p}^{\prime}$ at $\vec{r} 
^{\prime\prime}=\left( 0,0,-d\right) :$

\begin{equation*}
p^{\prime}=p\frac{\varepsilon_{m}-\varepsilon_{d}}{\varepsilon_{m}
	+\varepsilon_{d}}
\end{equation*}
\ with the result that 
\begin{equation}
\left\langle E_{z}\left( \vec{r}^{\prime}\right) E_{z}\left( \vec {r} 
^{\prime}\right) \right\rangle _{\omega}=\hbar\frac{\omega \varepsilon_{d} 
}{8\pi\sigma d^{3}}\coth\left( \frac{\hbar\omega}{2k_{B} T}\right) ,
\label{eq:ezimage}
\end{equation}
which is greater than $\left\langle E_{x}\left( \vec{r}^{\prime}\right)
E_{x}\left( \vec{r}^{\prime}\right) \right\rangle _{\omega}$ by a factor of
2. \ This anisotropy is quite significant for detailed exploration of the
theory by experiment.\ 

\paragraph{Induction regime}

This regime is characterized by the opposite limit $d\gg \delta.$ \ The qubit
is far away from the interface on the length scale of the penetration depth.
\ The image problem does not carry over directly since the electric field in
the metal satisfies $\nabla^{2}\vec{E}^{\left( f\right) }=\left( -2i/\delta
^{2}\right) \vec{E}^{\left( f\right) }$ in the metal and $\delta^{-2}$
cannot be neglected, as it was in the image regime. \ However, we may now
use the fact that the field penetrates only a short distance into the metal,
and this allows us to develop a perturbation series in $\omega$ for the
complex amplitudes $\vec{E}^{\left( f\right) },\vec{B}^{\left( f\right) }$
in the frequency domain. \ At order $\omega^{0}$ we have an electric field $ 
\vec {E}^{\left( f\right) }$ but $\vec{B}^{\left( f\right) }$ vanishes. \ $ 
\vec{E}^{\left( f\right) }$ is the static field from the previous image
calculation that is normal to the interface. \ At order $\omega^{1}$ there
is a magnetic field that corresponds to the static electric field according
to the equation $\nabla\times\vec{B}=-i\omega E/c.$ \ To compute $\vec {B} 
^{\left( f\right) }$ at this order we again put a dipole $\vec {p}=p\widehat{ 
	x}$ at $\vec{r}^{\prime}=\left( 0,0,d\right) $ together with its image
dipole $-p\widehat{x}$ at $\vec{r}^{\prime\prime}=\left( 0,0,-d\right) .$ \
This corresponds to a current $\vec{J}\left( \vec {r}\right) =p\left(
\partial~\delta^{3}\left( \vec{r}-\vec{r}^{\prime }\right) /\partial
x\right) -p\left( \partial~\delta^{3}\left( \vec {r}-\vec{r} 
^{\prime\prime}\right) /\partial x\right) .$ \ Computing the magnetic field
due to this current we have: 
\begin{equation*}
B_{y}^{\left( f\right) }\left( z=0\right) =\frac{-2ipd\omega}{c\left(
	\rho^{2}+d^{2}\right) ^{3/2}}\text{ and~}B_{z}^{\left( f\right) }\left(
z=0\right) =B_{x}^{\left( f\right) }\left(
z=0\right) = 0,
\end{equation*}
correct to order $\omega.$ \ $B_{y}^{\left( f\right) }$ is continuous at the
interface and $\nabla^{2}\vec{B}^{\left( f\right) }=-2i\delta^{-2}\vec {B} 
^{\left( f\right) }$ for $z<0.$ \ The crucial point is that since $ 
\delta^{-2}$ is large we may neglect the $x$ and $y$ derivatives in both $ 
\vec{B}^{\left( f\right) }$ and $\vec{E}^{\left( f\right) }$ for $z<0$ and
we have that 
\begin{equation*}
B_{y}^{\left( f\right) }=\frac{-2ipd\omega}{c\left( \rho^{2}+d^{2}\right)
	^{3/2}}\exp\left[ \left( 1-i\right) z/\delta\right] .
\end{equation*}
Since $\nabla\times\vec{E}=i\omega\vec{B}/c$ for $z<0$, consistency requires
that 
\begin{equation*}
\frac{\partial E_{x}^{\left( f\right) }}{\partial z}=\left( 1-i\right)
\delta^{-1}E_{x}^{\left( f\right) }\left( z\right) =i\omega B_{y}^{\left(
	f\right) }\left( z\right) /c
\end{equation*}
at order $\omega^{2}$. \ Solving these equations gives 
\begin{equation*}
E_{x}^{\left( f\right) }\left( z=0\right) =\frac{\left( 1+i\right)
	pd\delta\omega^{2}}{c^{2}\left( \rho^{2}+d^{2}\right) ^{3/2}}.
\end{equation*}
$E_{x}^{\left( f\right) }$ is continuous at the interface so we also get a
correction to the field for $z>0$ at order $\omega^{2}$. \ 

For $z>0$ the field components satisfy the Laplace equation $\nabla^{2}\vec { 
	E}^{\left( f\right) }=0,$ so we can get the field everywhere by applying
Green's theorem to the components of $\vec{E}^{\left( f\right) }$: 
\begin{equation*}
E_{x}^{\left( f\right) }( r) =-\frac{\left( 1+i\right)
	pd\delta\omega^{2}}{4\pi c^{2}}\int dx^{\prime}
dy^{\prime}\left( \rho^{\prime2}+d^{2}\right) ^{-3/2}\frac{\partial G_{D}}{ 
	\partial n^{\prime}}
\end{equation*}
where $G_{D}$ is the Dirichlet Green's function: $G_{D}=\left\vert \vec {r}- 
\vec{r}^{\prime}\right\vert ^{-1}-\left\vert \vec{r}-\vec{r}^{\prime
	\prime}\right\vert ^{-1}$ and $n^{\prime}$ is the outward-pointing normal.

Carrying out the integration and using Eq. (\ref{eq:enoise}) gives
\begin{align}
\left\langle E_{x}\left( \vec{r}\right) E_{x}\left( \vec{r}\right)
\right\rangle _{\omega} =\left\langle E_{y}\left( \vec{r}\right)E_{y}\left( \vec{r}\right) \right\rangle _{\omega} &=\frac{\hbar\omega}{8\pi d^{2}\sigma\delta}\coth\left( \frac {\hbar\omega}{2k_{B}T}\right) \label{eq:exinduction} \\
&\approx \begin{cases}
 \frac{\hbar\omega}{8\pi d^{2}\sigma\delta} & \text{ for } k_{B}T\ll \hbar\omega\text  \notag \\
\frac{\hbar\omega}{4\pi d^{2}\sigma\delta} & \text{ for } k_{B}T\gg \hbar\omega .  \notag
\end{cases}  \notag
\end{align}
Since $\delta \sim \frac{1}{\sqrt{\omega \sigma}}$, in the classical limit
we have that the noise is proportional to $\sqrt{\omega/\sigma},$ an
interesting contrast to the $\omega/\sigma$ dependence in the image regime.

For the $z$-$z$ correlation function the derivation is only slightly
different. \ We now put a dipole $\vec{p}=p\widehat{z}$ at $\vec{r}^{\prime
}=\left( 0,0,d\right) .$ \ $\vec{J}\left( \vec{r}\right) =p\left(
\partial~\delta^{3}\left( \vec{r}-\vec{r}^{\prime}\right) /\partial z\right)
+p\left( \partial~\delta^{3}\left( \vec{r}-\vec{r}^{\prime\prime }\right)
/\partial z\right) .$ \ The result for the noise spectral density is 
\begin{align}
\left\langle E_{z}\left( \vec{r},\omega\right) E_{z}\left( \vec{r} 
,\omega\right) \right\rangle _{\omega} & =\frac{\hbar\omega}{8\pi d^{2}\sigma\delta}\coth\left( \frac {	\hbar\omega}{2k_{B}T}\right) .
\label{eq:ezinduction}
\end{align}
This is the same as Eq. (\ref{eq:exinduction}), so the noise becomes isotropic
at large distances from a metal surface.

\paragraph{Summary of Approximate Results for the Point Qubit.}

The two regimes are distinguished by the relative magnitudes of $d$ and $ 
\delta$ - the distance of the source from the half space and the skin depth.
\ The following physical considerations serve as the basis for understanding
electric field noise in small devices. \ 

The image regime of small $d/\delta$ is fairly easily understood. \ In the
fictitious problem, the electric field penetrates the metal in the same way
it does in the textbook case of two dielectrics of strongly different
dielectric constants. \ The field is strongly screened at the surfaces so
that the field lines bend sharply at the interface. \ Thus the field in the
metal is almost parallel to the surface. \ This field dissipates energy at
the usual rate $\sim\sigma\left\vert \vec{E}\right\vert ^{2}$ per unit
volume in the fictitious problem, and the physical fluctuations are also
proportional to this. \ However, the ``impedance mismatch" dominates to the
extent that $\left\vert \vec{E}\right\vert \sim1/\sigma$ in the metal
overall and the noise spectral density at a given frequency is proportional
to $1/\sigma.$ \ The noise is stronger for poor conductors since the field
penetrates further. \ Once the dependence on the conductivity has been
determined, the $1/d^{3}$ spatial dependence follows by dimensional analysis
or noting that the fictitious field is produced by an image dipole.

The induction regime of large $d/\delta$ is somewhat different. \ The
electric field is essentially normal to the interface. \ This induces a
magnetic field parallel to the interface which penetrates only a distance $ 
\delta$ into the metal. \ This in turn induces an electric field that
dissipates energy. \ The volume in which the energy is dissipated is of
thickness $\delta$ rather than $d$, so the dissipation is proportional to $ 
\delta.$ \ Thus the image result is reduced by the factor $\delta/d$, and
the noise spectral density is proportional to $1/d^{2}\sqrt{\sigma}.$ \ 

\subsubsection{Extended Qubits}

For extended qubits, we need the full $\vec{r}$ and $\vec{r}^{\prime}$ dependence of $G.$ \ We compute using a method that will be used repeatedly in what follows. Details are given in the Appendix, along with explicit forms for the off-diagonal components of the noise tensors. We place a fictitious dipole $\vec{p} = p \hat{z}$ at $\vec{r}^{\prime}=\left( 0,0,d\right) $ and find the induced field 
\begin{align*}
\vec{E}^{\left( ind\right) }\left( \vec{r}\right) &=-\frac{p}{2\pi}\int
d^{2}q~\left( -iq_{x},-iq_{y},q\right) ~e^{-qd} \\ &\times \frac{1-\left(
	\varepsilon_{m}/\varepsilon_{d}\right) q/\alpha}{1+\left( \varepsilon
	_{m}/\varepsilon_{d}\right) q/\alpha}e^{i\vec{q}\cdot\vec{\rho}}e^{-qz}
\end{align*}
for $z > 0 $, and the corresponding electric noise is given by Eq. (\ref{eq:enoise}):
\begin{align}
&\left\langle \vec{E}\left( \vec{r}=\left( \vec{\rho},z\right) \right) E_{z}\left( \vec{r}^{\prime }=\left( 0,0,d\right) \right) \right\rangle_{\omega }=-\frac{\hbar }{2\pi }\coth \frac{\hbar \omega }{2k_{B}T} \notag  \\
&\times \Ima \int d^{2}q~\left( -iq_{x},-iq_{y},q\right) ~e^{-qd} \frac{1-\left(
	\varepsilon _{m}/\varepsilon _{d}\right) q/\alpha }{1+\left( \varepsilon
	_{m}/\varepsilon _{d}\right) q/\alpha }e^{i\vec{q}\cdot \vec{\rho}}e^{-qz}~
\label{eq:extEzz}
\end{align}

\begin{figure}[h]
	\centering
	\includegraphics[width=0.7\linewidth]{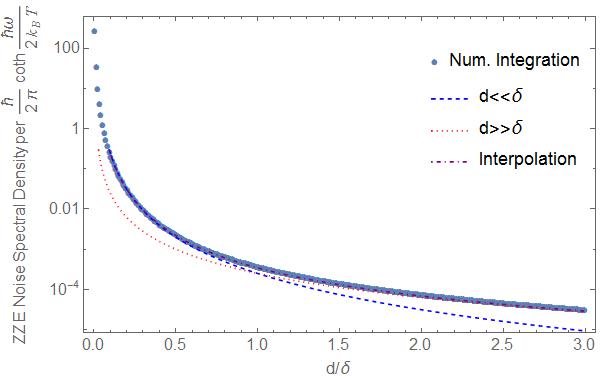}
	\caption{Numerical integration of Eq. (\protect\ref{eq:extEzz}), the
		electric noise spectral density for a localized qubit in the half-space
		geometry, compared with image and induction regime approximate results. $ 
		\frac{\protect\sigma}{\protect\omega} = 100$ and $p = 2.17$ for the
		interpolated function Eq. (\ref{eq:interp}). }
	\label{fig:halfspace_local_E_zz}
\end{figure}

The integral is complicated, but it can be evaluated numerically and it
simplifies in the limits of large and small $d$.

When $d\ll \delta,$ $\alpha\approx q$ and we find for the physical noise 
\begin{align}
&\left\langle \vec{E}\left( \vec{r} \right)
E_{z}\left( \vec{r}^{\prime} \right) \right\rangle
_{\omega} \approx-\frac{\hbar\omega\varepsilon_{d}}{2\pi\sigma}\coth\frac{\hbar\omega }{2k_{B}T} \nabla\frac{d+z}{\left[ \left( d+z\right) ^{2}+\rho^{2}\right] ^{3/2} 
}  \notag
\end{align}
The diagonal component is 
\begin{equation*}
\left\langle E_{z}\left( \vec{r} \right)
E_{z}\left( \vec{r}^{\prime} \right) \right\rangle
_{\omega}=\frac{\hbar\omega\varepsilon_{d}}{2\pi\sigma}\coth \frac{ 
	\hbar\omega}{2k_{B}T}\frac{2(d+z)^{2}
	-\rho^{2}}{\left[ \left( d+z\right) ^{2}+\rho^{2}\right] ^{5/2}},
\end{equation*}
and it can be verified that this equation reduces to Eq. (\ref{eq:ezimage}) when $\vec{r}=\vec{r}^{\prime}=\left( 0,0,d\right) ,$ an important check. \
This case has the unusual feature of anticorrelations in $E_{z}$ for large
lateral separations of $\vec{r}-\vec{r}^{\prime}$: $\rho>\sqrt{2}\left(
d+z\right) .$ \ This implies that in the appropriate geometry there can be
cancellations in the integral that determines qubit decoherence. \ This can
be incorporated as a design feature.\

For $d\gg\delta $ (but still $d\ll \delta \sigma /\omega )$ we have $\alpha
\approx \left( 1-i\right) \delta ^{-1}$ and the physical noise correlation
function is 
\begin{align*}
\left\langle \vec{E}\left( \vec{r}\right) E_{z}\left( \vec{r}^{\prime
}\right) \right\rangle _{\omega }& = -\frac{\hbar \omega \varepsilon _{d}}{2\pi \sigma \delta }\coth \frac{ 
\hbar \omega }{2k_{B}T}\nabla \frac{1}{\left[ \left( d+z\right) ^{2}+\rho
^{2}\right] ^{1/2}}
\end{align*} 
Specializing to the zz correlation function, 
\begin{equation*}
\left\langle E_{z}\left( \vec{r}\right) E_{z}\left( \vec{r}^{\prime }\right)
\right\rangle _{\omega }=\frac{\hbar \omega \varepsilon _{d}}{2\pi \sigma
	\delta }\coth \frac{\hbar \omega }{2k_{B}T}\frac{d+z}{\left[ \left(
	d+z\right) ^{2}+\rho ^{2}\right] ^{3/2}},
\end{equation*} 
and it can be verified that this equation reduces to Eq. \ref{eq:ezinduction}
when $\vec{r}=\vec{r}^{\prime }=\left( 0,0,d\right) .$ 
The situation for $\braket{\vec{E}(r) E_x(r')}_\omega $ is somewhat more complicated because of the lack of cylindrical symmetry. \ However, the method of the previous section does not depend on the symmetry and it can still be used. We now use $\vec{p} = p \hat{x}$. This leads to a fictitious induced electric field for $z>0:$ 
\begin{equation*}
\vec{E}^{\left( ind\right) }\left( \vec{r}\right) =-\frac{p}{2\pi }\frac{ 
	\partial }{\partial x}\nabla \int d^{2}q~\frac{1}{q}e^{-q\left( d+z\right) } 
\frac{1-\left( \varepsilon _{m}/\varepsilon _{d}\right) q/\alpha }{1+\left(
	\varepsilon _{m}/\varepsilon _{d}\right) q/\alpha }e^{iq_{x}x+iq_{y}y}
\end{equation*} 
and the physical noise correlation is 
\begin{equation*}
\left\langle \vec{E}\left( \vec{r}\right) E_{x}\left( \vec{r}^{\prime
}\right) \right\rangle _{\omega }=-\frac{\hbar }{2\pi }\frac{\partial }{ 
\partial x}\nabla \int d^{2}q~e^{-q\left( d+z\right) }\Ima \frac{ 
1-\left( \varepsilon _{m}/\varepsilon _{d}\right) q/\alpha }{1+\left(
\varepsilon _{m}/\varepsilon _{d}\right) q/\alpha }e^{iq_{x}x+iq_{y}y}\coth 
\frac{\hbar \omega }{2k_{B}T}.
\end{equation*}

For $d\ll \delta $ we have diagonal element of the physical noise spectral density
\begin{equation*}
\left\langle E_{x}\left( \vec{r}\right) E_{x}\left( \vec{r}^{\prime }\right)
\right\rangle _{\omega }=-\frac{\hbar }{2\pi }\frac{\omega \varepsilon _{d}}{ 
	\sigma }\frac{2x^{2}-\left( z+d\right) ^{2}-y^{2}}{\left[ \left( z+d\right)
	^{2}+\rho ^{2}\right] ^{5/2}}\coth \frac{\hbar \omega }{2k_{B}T},
\end{equation*} 
in agreement with Eq. (\ref{eximage}). \ At $\vec{r}=\vec{r}^{\prime }$ we
find 

\[\left\langle E_{x}\left( \vec{r}\right) E_{x}\left( \vec{r}=\vec{r}^{\prime}\right) \right\rangle _{\omega }=\frac{\hbar }{16\pi }\frac{\omega\varepsilon _{d}}{\sigma d^{3}}\coth \frac{\hbar \omega }{2k_{B}T}. \]
In the high $T$ limit this reduces to 
\[\left\langle E_{x}\left( \vec{r}\right) E_{x}\left( \vec{r}=\vec{r}^{\prime
}\right) \right\rangle _{\omega }=\frac{k_{B}T\varepsilon _{d}}{8\pi \sigma d^{3}}. \]

For $d\gg\delta $ (but still $d\ll \delta \omega /\sigma )$ we have
\begin{equation*}
\left\langle \vec{E}\left( \vec{r}\right) E_{x}\left( \vec{r}\right)
\right\rangle _{\omega }=\frac{\hbar }{2\pi }\frac{\omega \varepsilon _{d}}{ 
	\sigma \delta }\nabla \left\{ \frac{x}{\rho ^{2}}\left[ 1-\frac{d+z}{\left[
	\left( z+d\right) ^{2}+\rho ^{2}\right] ^{1/2}}\right] \right\} \coth \frac{ 
	\hbar \omega }{2k_{B}T},
\end{equation*} 
and the various components of the tensor may be calculated from this
expression. \ 

We have 
\begin{align*}
\left\langle E_{x}\left( \vec{r}\right) E_{x}\left( \vec{r}^{\prime }\right)
\right\rangle _{\omega} & =\frac{\hbar}{2\pi}\frac{\omega \varepsilon_{d}}{ 
	\sigma\delta}\coth\frac{\hbar\omega}{2k_{B}T}\times \\
& \left\{ \frac{y^{2}-x^{2}}{\rho^{4}}\left[ 1-\frac{d+z}{\left[ \left(
	z+d\right) ^{2}+\rho^{2}\right] ^{1/2}}\right] +\frac{x^{2}\left( d+z\right) 
}{\rho^{2}\left[ \left( z+d\right) ^{2}+\rho^{2}\right] ^{3/2}}\right\}
\end{align*}
and in particular at $\vec{r}=\vec{r}^{\prime}$ we find 
\begin{equation*}
\left\langle E_{x}\left( \vec{r}\right) E_{x}\left( \vec{r}\right) \right\rangle _{\omega}=\frac{\hbar\omega\varepsilon_{d}}{ 
	16\pi\sigma\delta d^{2}}\coth\frac{\hbar\omega}{2k_{B}T},
\end{equation*}
which is in agreement with Eq. (\ref{eq:exinduction}).

\subsubsection{Between Induction and Image Regimes}

In general Eq. (\ref{eq:extEzz}) cannot be simplified, but in both the image
and induction regime we can find analytic results (Eqs. (\ref{eq:ezimage})
and (\ref{eq:ezinduction})). Using these two results, we can interpolate a
function to compute correlation functions for qubit geometries that do not
fall into either of the extremal cases treated here. For two functions $f_1$
and $f_2$ we define a family of interpolated functions 
\begin{equation}
f_{int}(p) = \left(
f_1^p + f_2^p\right)^{\frac{1}{p}}
\label{eq:interp}
\end{equation} 
and search for the $p \in \mathbb{R}$
that optimizes the interpolated function's agreement with the extended qubit
noise spectral density. The interpolated functions are plotted alongside
numerical results for Eqs. (\ref{eq:extEzz}) and (\ref{eq:extBzz}) in Fig. 
\ref{fig:halfspace_local_E_zz} and Fig. \ref{fig:halfspace_local_B_zz}
respectively.
\subsection{Conducting Cylinder }

We consider a infinite conducting circular cylinder (conductivity $\sigma $
and radius $a)$ with its axis along the z-direction. \ There is a qubit at
the point $\vec{r}^{\prime }=\left( d,0,0\right) .$ \ We wish to compute $ 
\left\langle B_{i}\left( \vec{r}^{\prime }\right) B_{i}\left( \vec{r} 
^{\prime }\right) \right\rangle $ with $i=x,y,z.$ \ We're particularly
interested in the anisotropy of relaxation times, which depend on the ratios
of this correlation function for different values of $i.$ \ The most common
case is when the skin depth $\delta \gg a.$ \ We will also be mainly
interested in thin wires also in the sense that $d\ll  a.$ \ \ This means that
the fictitious applied field is slowly varying over the cylinder. \ The the
problem reduces to a computation of the electric polarizability.

The problem of the\textit{\ magnetic} polarizability of a conducting
cylinder in a uniform field is a standard one \cite{LLECM}. \ \
We modify the solution to obtain the electric polarizability $\vec{\beta}$,
defined by $P_{i}=\pi a^{2}\beta _{i}E_{i},$ where $P_{i}$ is the electric
dipole moment per unit length in direction $i.$ \ We find 
\begin{equation*}
\beta _{x}=\frac{1}{2\pi }\frac{4\pi i\sigma /\omega -C}{4\pi i\sigma
	/\omega +C}
\end{equation*} 
with 
\begin{equation*}
C=-1+\frac{kaJ_{0}\left( ka\right) }{J_{1}\left( ka\right) }.
\end{equation*} 
and  $k=\left( 1+i\right) /\delta .$ 

Again, the most interesting case is when $\delta \gg a,$ so $\left\vert
ka\right\vert \ll 1$ and 
\begin{equation*}
\frac{kaJ_{0}\left( ka\right) }{J_{1}\left( ka\right) }\approx ka\frac{1}{ 
	ka/2}=2,
\end{equation*} 
and then we find 
\begin{equation*}
\Ima \beta _{x}=\frac{\omega }{\pi \sigma }.
\end{equation*} 
When $d\gg a$ we can integrate along the z-axis assuming uniform applied
field. \ Some further details are given in Sec. 4. \ We find 

\[
	\left\langle E_{x}\left( \vec{r}^{\prime }\right) E_{x}\left( \vec{r} 
	^{\prime }\right) \right\rangle =\frac{123\omega \hbar a^{2}}{256\sigma d^{5}}\coth \left( \frac{\hbar
		\omega }{2k_{B}T}\right) 
\]
and
\[
	\left\langle E_{y}\left( \vec{r}^{\prime }\right) E_{y}\left( \vec{r} 
	^{\prime }\right) \right\rangle  =\frac{3\omega \hbar a^{2}}{32\sigma d^{5}}\coth \left( \frac{\hbar \omega 
	}{2k_{B}T}\right) .
\]
\ 

We may calculate the noise correlation for the z-direction in the same way.
\ However, this would seem to be problematic, since in a any finite wire the
electric flux would scome through the ends. \ We present the result as a
conjecture to be investigated in further work: 
\begin{equation*}
\left\langle E_{z}\left( \vec{r}^{\prime }\right) E_{z}\left( \vec{r} 
^{\prime }\right) \right\rangle =\frac{27\pi \hbar a^{4}}{2048d^{5}\delta
	^{2}}\coth \left( \frac{\hbar \omega }{2k_{B}T}\right) .
\end{equation*}

\bigskip
\subsection{Distant Object}

We now treat the electrical noise of a metallic object far away from the qubit ($d\gg L$). We consider a fictitious point dipole $\vec{p}$ at $\vec{r} ^{\prime}$, the metallic object approximated by a sphere at the origin and an observation point $\vec{r}$. Eq. (\ref{eq:efield}) gives the correlation function: 
\begin{align*}
&\left\langle E_{i}\left( \vec{r}\right) E_{k}\left( \vec{r}^{\prime }\right)
\right\rangle =\hbar\coth\left( \frac{\hbar\omega}{2k_{B}T}\right) \notag \\ 
& \times \Ima \left[
\alpha\left( \omega\right) \right] \frac{9x_{i}x_{k}^{\prime} ~\vec{r}\cdot 
	\vec{r}^{\prime}+\delta_{ik}r^{2}r^{\prime2}-3x_{i}x_{k}
	r^{\prime2}-3x_{i}^{\prime}x_{k}^{\prime}r^{2}}{r^{5}r^{\prime5}},
\end{align*}
where now $\vec{E}$ is the physical fluctuating field. \ The local noise at $ 
\vec{r}$ is 
\begin{equation}
\left\langle E_{i}\left( \vec{r}\right) E_{k}\left( \vec{r}^{\prime}=\vec{r} 
\right) \right\rangle =\hbar\coth\left( \frac{\hbar\omega}{2k_{B} T}\right) 
\Ima \left( \alpha\right) \frac{3x_{i}x_{k} +\delta_{ik}r^{2}}{r^{8}}.
\label{eq:polar}
\end{equation}

The $r^{-6}$ dependence is familiar from the van der Waals force, which has
a similar physical origin.\ \ 

The \textit{anisotropy} in lifetimes of a qubit in the presence of a
spherical electrode is independent of the value of $\alpha$. \ If the qubit
is located at $\vec{r}=r\widehat{z},$ then

\begin{align*}
\left\langle E_{x}\left( \vec{r}\right) E_{x}\left( \vec{r} 
\right) \right\rangle =\left\langle E_{y}\left( \vec{r}\right) E_{y}\left(\vec{r}\right) \right\rangle &=\hbar\coth\left( \frac{\hbar\omega}{2k_{B}T}\right) \frac{ \Ima \left[ \alpha\left(
	\omega\right) \right] }{r^{6}} \\
\left\langle  E_{z}(\vec{r}) E_z(\vec{r}) \right\rangle &=4\hbar\coth\left( \frac{\hbar\omega}{2k_{B} T}\right) \frac{\Ima \left[ \alpha\left( \omega\right) \right] }{r^{6}}.
\end{align*}
The anisotropy 
\begin{equation*}
\left\langle E_{z}\left( \vec{r}\right) E_{z}\left( \vec{r}\right)
	\right\rangle =4\left\langle E_{x}\left( \vec{r}\right) E_{x}\left( \vec {r} 
\right) \right\rangle 
\end{equation*}
is stronger than in the half-space case. Thus the problem of noise from a distant metallic object reduces to a
calculation of $\Ima \left[ \alpha\left( \omega\right) \right] ,$ the
dissipative part of the polarizability of the electrode. \ To get $\alpha,$
we need to calculate the change in the charge density of the electrode due
to a distant oscillating dipole, and the electric field that results from
this charge. We do this now in two limits.
\paragraph{Image Regime}

We first consider a metallic sphere of radius $a$ with $\delta\gg a.$ \ Once
again the fictitious problem is mathematically identical with that of a
dielectric sphere in a static field, so we may simply transcribe the
textbook formulas for the polarizability: 
\begin{equation}
\alpha=\frac{\varepsilon_{m}/\varepsilon_{d}-1}{\varepsilon_{m}/\varepsilon
	_{d}+2}a^{3}\approx\left( 1+\frac{3i\omega\varepsilon_{d}}{ 
	4\pi\sigma}\right) a^{3}.  \label{eq: esphere}
\end{equation}

Hence 
\begin{equation*}
\left\langle E_{i}\left( \vec{r}\right) E_{k}\left( \vec{r}\right)
\right\rangle =\frac{3\hbar\omega\varepsilon_{d}a^{3}}{4\pi\sigma}\frac { 
	3x_{i}x_{k}+\delta_{ik}r^{2}}{r^{8}}\coth\left( \frac{\hbar\omega}{2k_{B} T} 
\right)
\end{equation*}
For a metallic ellipsoid with radii $a_{x},a_{y},a_{z}$ in the $x,y,z$
directions the coordinate system is aligned with the axes of the ellipsoid
and the polarizability tensor satisfies $\alpha_{ij}=\delta_{ij}\alpha_{ii}$
with 
\begin{align*}
\alpha_{ii} & =\frac{1}{3}\frac{\varepsilon_{m}/\varepsilon_{d}-1}{1+\left(
	\varepsilon_{m}/\varepsilon_{d}-1\right) n_{i}}a_{x}a_{y}a_{z} \\
& \approx\left( 1+\frac{i\omega\varepsilon_{d}}{12n_{i}^{2}\pi\sigma}\right)
a_{x}a_{y}a_{z}.
\end{align*}
The depolarizing factors $n_{x},n_{y},n_{z}$ are positive and satisfy $n_{x}+n_{y}+n_{z}=1$ and \ $n_{i}$ are decreasing functions of $a_{i}.$ \ In particular, if $ 
a_{x}<a_{y}<a_{z}$ then $n_{x}>n_{y}>n_{z}.$ The connection between the $n_{i}$ and the $a_{i}$ involves elliptic
integrals. \ Exact expressions and tables may be found in \cite{Osborn}. Using Eq. (\ref{eq:eiekanis}) we have 
\begin{align*}
\left\langle E_{i}\left( \vec{r}\right) E_{k}\left( \vec{r}^{\prime }\right)
\right\rangle & =\hbar\coth\left( \frac{\hbar\omega}{2k_{B}T}\right) \Ima \left( \alpha_{jj}\right) f_{kj}\left( \vec {r}^{\prime}\right)
f_{ij}\left( \vec{r}\right) \\
& =\frac{\hbar\omega \varepsilon_d V}{16\pi^2\sigma}\coth\left( \frac { 
	\hbar\omega}{2k_{B}T}\right) \times\frac{1}{n_{j}^{2}}\frac{3x_{k}^{\prime
	}x_{j}^{\prime}-\delta_{kj}r^{\prime2}}{r^{\prime5}}\frac{3x_{i}x_{j}
	-\delta_{ij}r^{2}}{r^{5}},
\end{align*}
a distance $r$ from the center of the ellipsoid of volume $V$.\ To understand the physics of
this formula, think of a qubit at $\vec{r}=r\widehat{z}$ with the origin of coordinates at the center of the ellipsoid. \ Then the off-diagonal components of the noise tensor vanish and the formula exhibits the anisotropy mentioned above. This expression confirms the intuition that the noise should be stronger in the directions where the axis
is longer, since the polarizability is greater. \

\paragraph{Induction regime}

\ Again we first consider a metallic sphere of radius $a.$ \ The electric
field outside the sphere in lowest order in $\omega$ in spherical
coordinates is 
\begin{equation*}
\vec{E}^{\left( 0\right) }=\widehat{r}E_{0}\left( 1+\frac{2a^{3}}{r^{3}} 
\right) \cos\theta-\widehat{\theta}E_{0}\left( 1-\frac{a^{3}}{r^{3}}\right)
\sin\theta
\end{equation*}
and solving the equation $\nabla\times\vec{B}^{\left( 0\right)
}=i\lambda^{-1}\vec{E}^{\left( 0\right) }$ we get a corresponding magnetic
field 
\begin{equation*}
\vec{B}_{out}^{\left( 0\right) }=-\frac{iE_{0}}{2\lambda}\left( r+\frac{ 
	2a^{3}}{r^{2}}\right) \sin\theta~\widehat{\phi}
\end{equation*}
and using this as a boundary condition for the solution of the diffusion
equation inside the sphere gives 
\begin{equation*}
\vec{B}_{in}^{\left( 0\right) }=-\frac{3iE_{0}a}{2\lambda}\sin
\theta~e^{-\left( 1-i\right) \left( a-r\right) /\delta}~\widehat{\phi}
\end{equation*}
where again the condition $\delta\ll a$ has been used to neglect the
tangential derivatives. \ Since $\vec{E}_{in}=\left( c/4\pi\sigma\right)
\nabla \times\vec{B}_{in}$ this gives an electric field 
\begin{align*}
\vec{E}_{in}^{\left( 1\right) } & =-\widehat{\theta}\left( c/4\pi
\sigma\right) \left( -\frac{3iE_{0}a}{2\lambda}\right) \sin\theta\frac {1}{r} 
\left[ \frac{\partial}{\partial r}re^{-\left( 1-i\right) \left( a-r\right)
	/\delta}\right] \\
& =\widehat{\theta}\left( \frac{3\left( 1+i\right) E_{0}\omega a}{ 
	8\pi\sigma\delta}\right) \sin\theta~e^{-\left( 1-i\right) \left( a-r\right)
	/\delta},
\end{align*}
and since this field is tangential it is continuous at the boundary we can simply compare this field with the field of a dipole in the $z$-direction: $\vec{E}_{dip}=\left( p\sin\theta/a^{3}\right) \widehat {\theta}$ 
, together with $p=\alpha E_{0}$ and we find 
\begin{equation*}
\alpha=\frac{3\left( 1+i\right) \omega a^{4}}{8\pi\delta\sigma}
\end{equation*}
for the polarizability in the induction regime. \ 
Using Eq. (\ref{eq:polar}), we have that if the qubit is located at $\vec {r} 
=r\widehat{z},$ then
\begin{equation*}
\left\langle E_{x}\left( \vec{r}=r\widehat{z}\right) E_{x}\left( \vec {r}=r 
\widehat{z}\right) \right\rangle =\left\langle E_{y}\left( \vec {r}=r 
\widehat{z}\right) E_{y}\left( \vec{r}=r\widehat{z}\right) \right\rangle = 
\frac{3\hbar\omega a^{4}~}{8\pi\sigma\delta r^{6}}\coth\left( \frac{ 
	\hbar\omega}{2k_{B}T}\right)
\end{equation*}

\begin{equation*}
\left\langle E_{z}\left( \vec{r}=r\widehat{z}\right) E_{z}\left( \vec {r}=r 
\widehat{z}\right) \right\rangle =\frac{3\hbar\omega a^{4}~}{2\pi
	\sigma\delta r^{6}}\coth\left( \frac{\hbar\omega}{2k_{B}T}\right) .
\end{equation*}

\paragraph{General result}

The problem of the polarization of a metallic sphere is exactly solvable for
all $d/\delta$ but it is not trivial. \ The method may be found in \cite 
{Garg}, and it is discussed in \cite{Bookgarg}, but seems not to have been
solved previously!

The polarizability $\alpha$ for the sphere of radius $a$ is given by 
\begin{equation*}
a^{-3}\alpha=-\frac{1}{2}\frac{\kappa^{2}a~j_{0}+\left( 1+2\varepsilon
	\right) ~j_{0}^{\prime}}{\kappa^{2}a~j_{0}+\left( 1-\varepsilon\right)
	~j_{0}^{\prime}}.
\end{equation*}
The symbols are defined as $\kappa=\left( 1+i\right) /\delta,$ $j_{0}=\left(
1/\kappa a\right) \sin\kappa a,$ $j_{0}^{\prime}=\left( 1/a\right)
\cos\kappa a-\left( 1/\kappa a^{2}\right) \sin\kappa a.$

To obtain the first correction in the case $\delta\gg a$ we expand to first
order in $\omega/\sigma$ and $a/\delta,$ (always assuming $\omega
/\sigma\ll a/\delta)$ and find 
\begin{equation*}
j_{0}\approx1,~j_{0}^{\prime}\approx-\frac{1}{3}\kappa^{2}a
\end{equation*}
and we have 
\begin{equation*}
\alpha=a^{3}\left( 1+\frac{3i\omega}{4\pi\sigma}\right) ,
\end{equation*}
in agreement with Eq. (\ref{eq: esphere}) for the dissipative part. \ Note
that the term that is zeroth-order in $\omega$ gives a polarizability $ 
\alpha =a^{3},$ which is the proper static limit given in many textbooks.

When $\delta\ll a,$ then 
\[
j_{0}  \approx\frac{i}{2\kappa a}e^{-i\left( 1+i\right) a/\delta},\; \; \; 
j_{0}^{\prime} \approx\frac{1}{2a}e^{-i\left( 1+i\right) a/\delta}
\]
and
\[\alpha \approx a^{3}\left( 1+\frac{3a\left( 1+i\right) \omega~}{8\pi\sigma \delta} 
\right) .\]
As $\omega$ increases, we find that $\Ima \alpha$ increases, so it is a
monotonic function of $\omega$.

\subsection{Multiple Objects}

Real devices tend to have complex geometries with multiple metallic device
elements. \ A modern spin qubit experiment may involve a back gate or an
accumulation gate having a layer or half-space shape. \ There may be up to
tens of finger gates for lateral or voltage control that are approximately
cylindrical. \ Clearly a numerical approach is indicated for these cases,
which is beyond the scope of this paper. \ We therefore limit ourselves to a
few remarks. \ 

In many cases, it may be reasonable to regard different metallic elements as
noise sources that are statistically independent. \ If this assumption
holds, then 
\begin{equation}
\left\langle E_{i}\left( \vec{r}\right) E_{j}\left( \vec{r}^{\prime }\right)
\right\rangle _{\omega}=\sum\limits_{s=1}^{N}\left\langle E_{i}\left( \vec{r} 
\right) E_{j}\left( \vec{r}^{\prime}\right) \right\rangle _{\omega}^{\left(
	s\right) },  \label{eq:inc}
\end{equation}
where the $\left( s\right) $ indexes the sources, of which there are $N$
total. The various noise sources add incoherently.

The physical analogy of Sec. 2 shows that this assumption cannot be be
strictly correct. The various device elements are in fact all driven by a
single fictitious dipole and they are therefore in phase. However, unless
the qubit occupies a position of high symmetry with regard to at least one
pair of metallic objects. This can occur: it is common to place qubits near
the tips of opposing finger gates. \ However, in most other cases the
symmetry is low and Eq. (\ref{eq:inc}) can be used. 

\subsection{Sharp Points}
A serious concern for qubit decoherence is the geometrical enhancement of
noise in the neighborhood of surface asperities of conductors. \ The
question is whether the well-known divergence of local field strengths at
such structures carries over to noise. \ This is a particularly pressing
issue for for semiconductor qubits where finger gates with sharp points are
a standard feature of device architectures. \ 
\begin{figure}[tbp]
	\centering
	\includegraphics[width=0.7\linewidth]{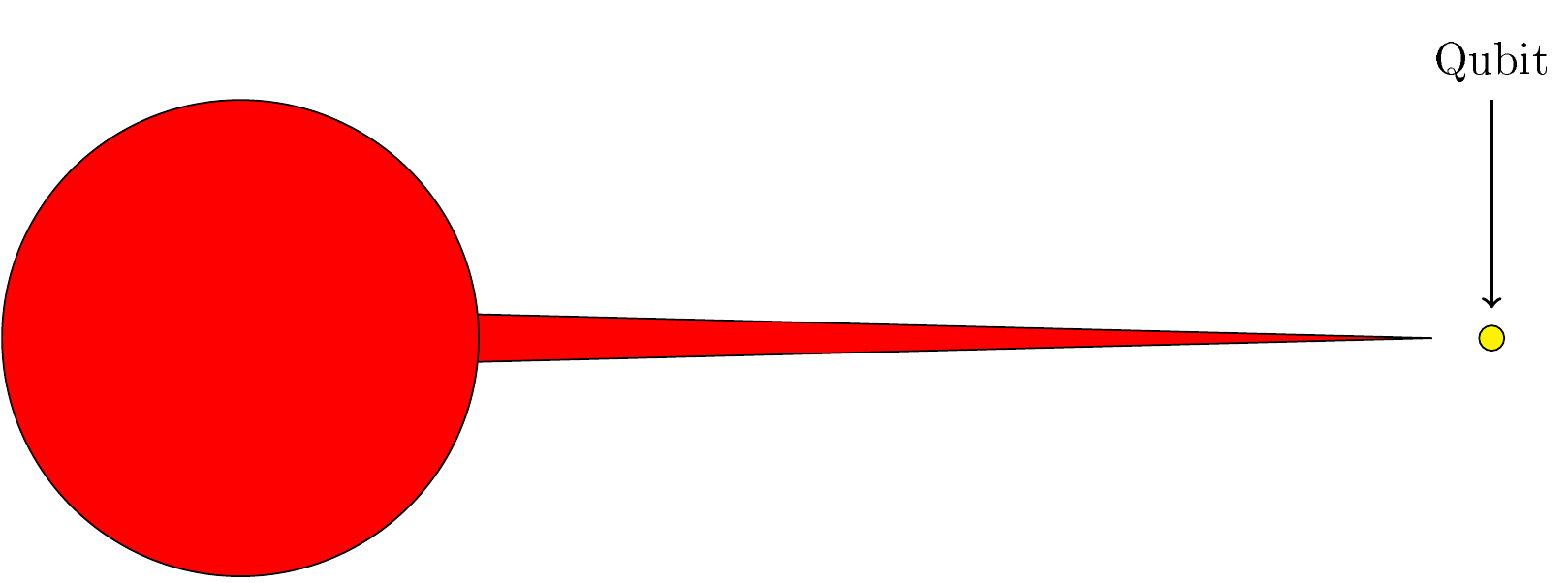}
	\caption{}
	\label{fig:sph}
\end{figure}
However, it can be seen fairly simply that electric noise is not greatly
enhanced by asperities in the case that $\delta$ is greater than the size of
the surface feature (the usual case). \ We imagine a spherical geometry with
a sharp point added on top, and a qubit near the point \ (See Fig. \ref{fig:sph}).
\ Qualitatively, the quasistatic
electric field lines will gather at the point, giving the familiar
lightning-rod effect. \ However, these lines are outside the object and they
do not produce the dissipation that is associated with field fluctuations
and noise. \ Inside, the magnitude of the field is reduced by a factor $ 
\omega/\sigma.$ \ This internal field produces currents and it is therefore
the field associated with the dissipative part of the response, and, in
turn, to the noise strength. \ These currents run away from the point and
give rise to the surface charge\ $\sigma\left( \vec{r}^{\prime }\right) $
whose density diverges at the tip as $r^{-1+\nu}$, $\nu=\left[ 2\ln\left(
2/\alpha\right) \right] ^{-1},$ where $\alpha$ is half the opening angle of
the tip, taken here as a cone (\textit{i.e.,} $\alpha=0$ for an infinitely
sharp tip). \ The presence of the logarithm means that $\nu\ll 1$ even for a
very sharp tip. \ Note $\nu>0. $\ 

The z component of the fictitious electric field at the point $\vec{r} 
=\left( 0,0,-d\right) $ is proportional to 
\begin{align*}
E_{z}^{\left( f\right) }\left( \vec{r}\right) &
\sim\int_{0}^{p}d^{2}r^{\prime}\frac{\sigma\left( \vec{r}^{\prime}\right)
	\left( \vec {r}-\vec{r}^{\prime}\right) _{z}}{\left\vert \vec{r}-\vec{r} 
	^{\prime }\right\vert ^{3}} \\
& \sim\int_{0}^{p}\sin\alpha~r^{\prime}dr^{\prime}\frac{\left( r^{\prime
	}\right) ^{-1+\nu}\left( -d-r^{\prime}\cos\alpha\right) }{\left\vert \left(
	r^{\prime}\right) ^{2}+d^{2}+2dr^{\prime}\cos\alpha\right\vert ^{3/2}}
\end{align*}
where $p$ is an upper cutoff on the size of the cone. \ We are only
interested in the small $d$ behavior, which follows from simple scaling
arguments as $E_{z}^{\left( f\right) }\sim d^{\nu+1/2}$ and this carries
over to the physical field fluctuations $\left\langle E_{z}\left( \vec{r} 
\right) E_{z}\left( \vec{r}\right) \right\rangle _{\omega}\sim d^{\nu+1/2}. $
\ So the divergence of the fields as the point is approached along the
surface does not carry over to the noise in the immediate region near the
tip but outside the conductor. \ \ \ \ \ 
\subsection{Charge qubits}

To understand qubit decoherence in the presence of noise, the frequency
dependence of the noise is of paramount importance. \ To this end, write the
noise spectral density $\left\langle E_{i}\left( \vec{r}\right) E_{j}\left( 
\vec{r}^{\prime}\right) \right\rangle $from EWJN as 
\begin{equation*}
\left\langle E_{i}\left( \vec{r}\right) E_{j}\left( \vec{r}^{\prime }\right)
\right\rangle _{\omega}=f\left( \omega\right) ~\omega~\coth\left( \frac{ 
\hbar\omega}{2k_{B}T}\right) ,
\end{equation*}
where all spatial and device geometry information is contained in $f\left(
\omega\right) .$ \ For EWJN, $f\left( \omega\right) \rightarrow$ $f_{0}$, a
constant as $\omega\rightarrow0.$ \ $f_{0}$ sets the overall scale of the
noise strength. \ In addition there is a high-frequency cutoff $1/\tau$ at
the relaxation time for the conduction electrons in the metal. \ Thus $ 
f\left( \omega\right) \rightarrow0$ when $\omega\gg 1/\tau.$ \ Physically, the
factor of $\omega$ comes from the connection of noise to dissipation. \
Photons are non-interacting bosons - hence the cotangent factor. \ \ This
sort of noise is white, or at least white-ish. \ This means that echo
techniques are not likely to be very useful for extending qubit lifetimes
when EWJN is the dominant source of decoherence. \ This noise is essentially
the same as that of the well-known spin-boson model and the results are well
known, so we only briefly summarize results here and give no derivations. \ 

There are three frequency regime for the spectral density. \ 1. When $ 
0\leq\omega<2k_{B}T/\hbar$,$~$then $\left\langle E_{i}\left( \vec{r}\right)
E_{j}\left( \vec{r}^{\prime}\right) \right\rangle
_{\omega}=2k_{B}Tf_{0}/\hbar.$ \ 2. When $2k_{B}T/\hbar<\omega<1/\tau$ we
have $\left\langle E_{i}\left( \vec{r}\right) E_{j}\left( \vec{r} 
^{\prime}\right) \right\rangle =f\left( \omega\right) ~\omega,$ where
typically the frequency dependence of $f\left( \omega\right) $ is weak. \ 3.
When $\omega>1/\tau,$ then the frequency dependence is material-dependent
but we may usually assume that the noise is cut off. \ 

In regime 1, the fluctuations are thermal. \ Regime 2 is the quantum regime
and the linear spectrum is refered to as ``ohmic". \ Regime 3 is above the
the high-frequency cutoff, whose presence is implicit in this paper. \ The
symbol $\sigma$ denotes the DC conductivity; however, no equation in which
it appears can be used at frequencies greater than $1/\tau.$ This frequency
range is generally in the infrared for metals. \ 

The qubit energy level separation is $\hbar\omega_{0}$ and $\omega_{0}$ may
be in either Regime 1 or Regime 2, depending on the implementation. \ No
existing implementation operates in Regime 3. \ \

\section{Magnetic Noise}
\subsection{Half Space\ }

For magnetic noise the image method is not useful, so we proceed directly to
general results for extended qubits.

\subsubsection{Extended Qubits}

\ \ \ Again, we are interested in a metal with conductivity $\sigma$ that
occupies the half space $z<0.$ \ The equations satisfied by the fields are
the same as for the electric case. \ The only difference for magnetic fields
is that $\vec{B},$ unlike $E,$ is continuous at the interface, since we
dealing with non-magnetic materials.

For this problem we place a fictitious magnetic dipole moment $\vec{m}$ at
the point $\vec{r}^{\prime}=\left( 0,0,d\right) $ 
and the physical noise spectral density is 
\begin{figure}[h!]
\centering
\includegraphics[width=0.7\linewidth]{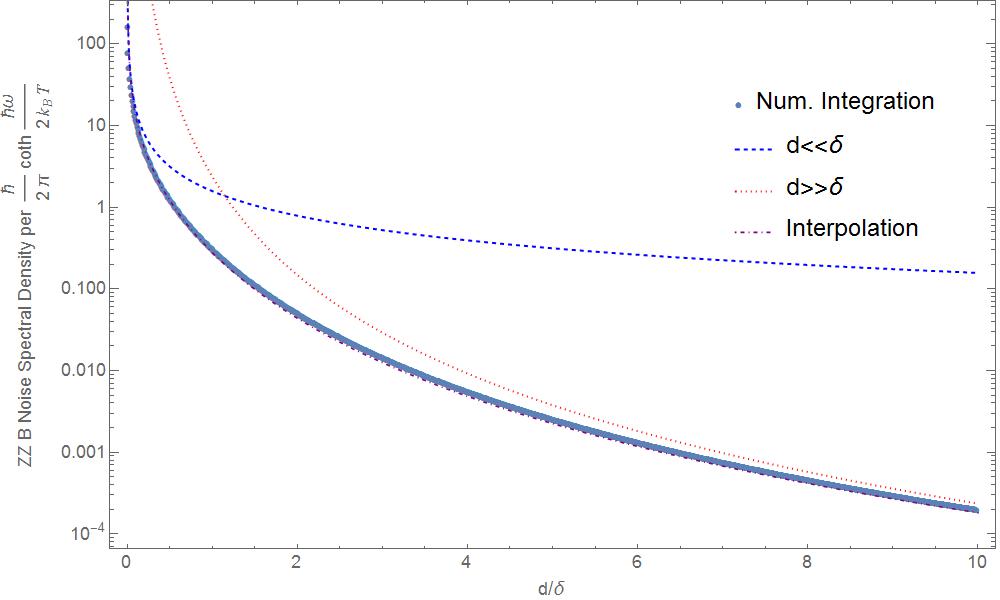}
\caption{Numerical integration of Eq. (\protect\ref{eq:extBzz}), the
magnetic noise spectral density for a localized qubit in the half-space
geometry, compared with image and induction regime approximate results. $ 
\frac{\protect\sigma}{\protect\omega} = 100$ and $p = -0.358$ for the
interpolated function Eq. (\ref{eq:interp}). }
\label{fig:halfspace_local_B_zz}
\end{figure}
\begin{align}
\left\langle \vec{B}\left( \vec{r}\right) B_{z}\left( \vec{r}^{\prime
}\right) \right\rangle _{\omega} & =\frac{\hbar}{4\pi\delta^{2}}\coth \frac{ 
\hbar\omega}{2k_{B}T}\Ima \int d^{2}q\frac{1}{q^{2}}~\left(
q_{x},q_{y},iq\right) e^{-q\left( z+d\right) }e^{iq_{x}x+iq_{y}y}
\label{eq:extBzz} \\
& =-\frac{\hbar}{4\pi\delta^{2}}\coth\frac{\hbar\omega}{2k_{B}T}\nabla\int
d^{2}q\frac{1}{q^{2}}e^{-q\left( z+d\right) }e^{iq_{x}x+iq_{y}y}  \notag
\end{align}

The diagonal element is 
\begin{align*}
\left\langle B_{z}\left( \vec{r}\right) B_{z}\left( \vec{r}^{\prime }\right)
\right\rangle _{\omega} & =\frac{\hbar}{4\pi\delta^{2}}\coth \frac{ 
\hbar\omega}{2k_{B}T}\int d^{2}q\frac{1}{q}e^{-q\left( z+d\right)
}e^{iq_{x}x+iq_{y}y} \\
& =\frac{\hbar}{2\delta^{2}}\frac{1}{\left[ \left( d+z\right) ^{2}+\rho ^{2} 
\right] ^{1/2}}\coth\frac{\hbar\omega}{2k_{B}T},
\end{align*}
and off-diagonal components are in the appendix.

For a point qubit $(\vec{r} = \vec{r}^{\prime })$ we have
\[
\left\langle B_{z}\left( \vec{r}\right) B_{z}\left( \vec {r} \right) \right\rangle _{\omega}  =\frac{\hbar}{4d\delta^{2}}\coth 
\frac{\hbar\omega}{2k_{B}T}. \]
And in the high $T$ limit this reduces to
\[\left\langle B_{z}\left( \vec{r}\right) B_{z}\left( \vec {r} \right) \right\rangle _{\omega} = \frac{k_{B}T}{2\omega d\delta^{2}}.\]

For $d\gg \delta$ we have 
\begin{equation*}
\frac{1-q/\alpha}{1+q/\alpha}\approx\left( 1+q\delta+iq\delta\right) ,
\end{equation*}
and 
\begin{align*}
\left\langle \vec{B}\left( \vec{r}\right) B_{z}\left( \vec{r}^{\prime}\right) \right\rangle _{\omega} 
&=-\frac{\hbar\delta}{2\pi}\coth\frac{\hbar\omega}{2k_{B}T}\nabla\int d^{2}q~q~e^{-q\left( z+d\right) }e^{iq_{x}x+iq_{y}y}
\end{align*}
The diagonal component is 
\begin{equation*}
\left\langle B_{z}\left( \vec{r}\right) B_{z}\left( \vec{r}^{\prime }\right)
\right\rangle _{\omega}=-\hbar\delta\coth\frac{\hbar\omega}{2k_{B}T}\times 
\left[ \frac{-6\left( z+d\right) ^{3}+9\rho^{2}\left( z+d\right) }{\left[
\left( d+z\right) ^{2}+\rho^{2}\right] ^{7/2}}\right] ,
\end{equation*}
For a point qubit only the diagonal component is nonzero: 
\begin{equation*}
\left\langle B_{z}\left( \vec{r}\right) B_{z}\left( \vec {r}\right) \right\rangle _{\omega}=\frac{3\hbar\delta}{8d^{4}}\coth 
\frac{\hbar\omega}{2k_{B}T}.
\end{equation*}

\subsubsection{Solution for $\vec{m}=m\widehat{x}.$}

This is more complicated because of the lack of cylindrical symmetry, but
the essential procedure is the same. \ We find 
\begin{align*}
\left\langle \vec{B}\left( \vec{r}\right) B_{x}\left( \vec{r}^{\prime
}\right) \right\rangle _{\omega} &=\frac{\hbar}{2\pi}\coth\frac{\hbar\omega}{2k_{B}T}\frac{\partial}{ 
\partial x}\nabla\Ima \int d^{2}q~\frac{1}{q}\frac{1-q/\alpha}{ 
1+q/\alpha }e^{iq_{x}x+iq_{y}y-q\left( z+d\right) }~
\end{align*}
For $d\ll \delta$ this is 
\begin{align*}
\left\langle \vec{B}\left( \vec{r}\right) B_{x}\left( \vec{r}^{\prime
}\right) \right\rangle _{\omega} 
&=-\frac{\hbar}{2\delta^{2}}\coth\frac{\hbar\omega}{2k_{B}T}\frac{\partial 
}{\partial x}\nabla\int_{0}^{\infty}dq~\frac{1}{q^{2}}e^{-q\left( z+d\right)
}J_{0}\left( q\rho\right)
\end{align*}
The diagonal component is 
\begin{align*}
\left\langle B_{x}\left( \vec{r}\right) B_{x}\left( \vec{r}^{\prime }\right)
\right\rangle _{\omega} 
&=\frac{\hbar}{2\delta^{2}}\coth\frac{\hbar\omega}{2k_{B}T}\left[ \frac{ 
x^{2}-y^{2}}{\rho^{2}}\left\{ \left[ \left( d+z\right) ^{2}+\rho ^{2}\right]
^{1/2}-\left( d+z\right) \right\} -\frac{x^{2}}{\rho ^{2}\left[ \left(
d+z\right) ^{2}+\rho^{2}\right] ^{1/2}}\right].
\end{align*}
At $\vec{r}=\vec{r}^{\prime}$ we have that only the diagonal component is
non-vanishing and 
\[
\left\langle B_{x}\left( \vec{r}^{\prime}\right) B_{x}\left( \vec {r} 
^{\prime}\right) \right\rangle =\frac{\hbar}{8d\delta^{2}}\coth \frac{ 
\hbar\omega}{2k_{B}T}.\]
In the high $T$ limit this reduces to 
\[\left\langle B_{x}\left( \vec{r}^{\prime}\right) B_{x}\left( \vec {r} 
^{\prime}\right) \right\rangle =\frac{k_{B}T}{4d\omega\delta^{2}}.\]
For $d\gg \delta$ we get 
\begin{align*}
\left\langle \vec{B}\left( \vec{r}\right) B_{x}\left( \vec{r}^{\prime
}\right) \right\rangle _{\omega} 
& =-\hbar\delta\coth\frac{\hbar\omega}{2k_{B}T}\frac{\partial}{\partial x} 
\nabla\frac{\partial}{\partial z}\frac{1}{\left[ \left( d+z\right)
^{2}+\rho^{2}\right] ^{1/2}}.
\end{align*}
This yields

\begin{equation*}
\left\langle B_{x}\left( \vec{r}\right) B_{x}\left( \vec{r}^{\prime }\right)
\right\rangle _{\omega}=\hbar\delta\coth\frac{\hbar\omega}{2k_{B}T}\left(
d+z\right) \frac{3\left( d+z\right) ^{2}+3\rho^{2}-15x^{2}}{\left[ \left(
d+z\right) ^{2}+\rho^{2}\right] ^{7/2}}
\end{equation*}
For $\vec{r}=\vec{r}^{\prime}$ only the diagonal component is nonzero: 
\[
\left\langle B_{x}\left( \vec{r}^{\prime}\right) B_{x}\left( \vec {r} 
^{\prime}\right) \right\rangle _{\omega}  =\frac{3\hbar\delta}{16\pi d^{4}} 
\coth\frac{\hbar\omega}{2k_{B}T}. \]
In the high $T$ limit this becomes
\[\left\langle B_{x}\left( \vec{r}^{\prime}\right) B_{x}\left( \vec {r} 
^{\prime}\right) \right\rangle _{\omega}=\frac{3\hbar\delta}{8\pi d^{4}}.\]

\subsection{Cylinder}

In this section, we consider a infinitely long conducting circular cylinder
as a source of EWJN. \ The cylinder has conductivity $\sigma$ and radius $a$
and its axis is along the z-direction. This geometry is important, since
cylindrical microwave antennas are used for single qubit rotations. \ \
There is a qubit at the point $\vec{r}'=\left( d,0,0\right) .$ \ We wish to
compute $\left\langle B_{i}\left( \vec{r}'\right) B_{i}\left( \vec {r}'\right)
\right\rangle $ with $i=x,y,z.$ \ We're particularly interested in the
anisotropy of relaxation times, which depend on the ratios of this
correlation function for different values of $i.$ \ The most common case is
when the skin depth $\delta\gg a.$ \ We will also be mainly interested in thin
wires also in the sense that $d\gg a.$

We need the solution to the problem of the magnetic polarizability of a
conducting cylinder in a uniform field. \ This is given by \cite{LLECM}. \ \
The polarizabilities $\alpha_{i}$ are defined by the formulas 
\begin{align*}
M_{x} & =\pi a^{2}\alpha_{x}B_{x} \\
M_{y} & =\pi a^{2}\alpha_{y}B_{y} \\
M_{z} & =\pi a^{2}\alpha_{z}B_{z},
\end{align*}
where $M_{i}$ is the magnetic moment per unit length in direction $i.$ \ Here 
\begin{align*}
\alpha_{x} & =\alpha_{y}=-\frac{1}{2\pi}\left[ 1-\frac{2}{ka}\frac { 
J_{1}\left( ka\right) }{J_{0}\left( ka\right) }\right] , \\
\alpha_{z} & =-\frac{1}{4\pi}\left[ 1-\frac{2}{ka}\frac{J_{1}\left(
ka\right) }{J_{0}\left( ka\right) }\right] ,
\end{align*}
with $k=\left( 1+i\right) /\delta.$ \ We will mainly need the imaginary part
in the limit where $\delta\gg a,$ which is 
\begin{align*}
\Ima \alpha_{x} & =\Ima \alpha_{y}=\frac{a^{2}}{8\pi\delta^{2}} \\
\Ima \alpha_{z} & =\frac{a^{2}}{16\pi\delta^{2}}.
\end{align*}

For illustration purposes we give some details of the calculation for $i=j=z$ 
.

The fictitous dipole $\vec{m}=m\widehat{z}$ at the point $\vec{r}^{\prime }=(d,0,0)$ 
sets up a field 
\begin{equation*}
B_{z}\left( \vec{r}\right) =m\frac{3\left( \vec{r}-\vec{r}^{\prime}\right)
_{z}\left( \vec{r}-\vec{r}^{\prime}\right) _{z}-\delta_{zz}\left\vert \vec{r} 
-\vec{r}^{\prime }\right\vert ^{2}}{\left\vert \vec{r}-\vec{r}^{\prime
}\right\vert ^{5}}
\end{equation*}
and along the axis of the cylinder this is \ \ 
\begin{equation*}
B_{z}\left( \vec{r}=\left( 0,0,z\right) \right) =m\frac{2z^{2}-d^{2}}{\left(
z^{2}+d^{2}\right) ^{5/2}}.
\end{equation*}
Note that this changes sign as a function of $z,$ which is a characteristic
of the dipole field in this geometry. \ This field will set up currents in
the cylinder that give rise to the induced field. \ If $d\gg a,$ we may do
WKB: use the uniform solution but with an applied field that varies slowly
with $z.$ \ In this approximation the dipole moment per unit length at $z$
is proportional to the applied field at $z$ with the polarizability already
given above. \ Thus 
\begin{align*}
M_{z}\left( z\right) & =\pi a^{2}\alpha_{z}B\left( z\right) \\
& =\pi a^{2}\alpha_{z}m\frac{2z^{2}-d^{2}}{\left( z^{2}+d^{2}\right) ^{5/2}}
\end{align*}
and this sets up an induced field at $\vec{r}$ which is 
\begin{align*}
B_{z}^{\left( ind\right) } & =\int_{-\infty}^{\infty}dz~M_{z}\left( z\right) 
\frac{2z^{2}-d^{2}}{\left( z^{2}+d^{2}\right) ^{5/2}} \\
& =\frac{\pi a^{2}\alpha_{z}m}{d^{5}}I_{z},
\end{align*}
where 
\begin{equation*}
I_{z}=\int_{-\infty}^{\infty}dx\frac{\left( 2x^{2}-1\right) ^{2}}{\left(
1+x^{2}\right) ^{5}}=\frac{27\pi}{128.}.
\end{equation*}
According to the usual prescription, then we find 
\begin{equation*}
\left\langle B_{z}\left( \vec{r}\right) B_{z}\left( \vec{r}\right)
\right\rangle =\frac{27\pi\hbar a^{2}}{256d^{5}}\Ima \left[ \frac{2}{ka} 
\frac{J_{1}\left( ka\right) }{J_{0}\left( ka\right) }\right] \coth\left( 
\frac{\hbar\omega}{2k_{B}T}\right) ,
\end{equation*}
valid for any value of $\delta/a.$

When $a\gg \delta$ we expand the Bessel functions for small argument and find 
\begin{align*}
B_{z}^{\left( ind\right) } & =\frac{\pi a^{2}\alpha_{z}m}{d^{5}}\frac {27\pi 
}{128} \\
\left\langle B_{z}\left( \vec{r}\right) B_{z}\left( \vec{r}\right)
\right\rangle & =\frac{27\pi\hbar a^{4}}{2048d^{5}\delta^{2}}\coth\left( 
\frac{\hbar\omega}{2k_{B}T}\right) .
\end{align*}

The same computation can be performed for the x and y directions. \ The
results for $i=j=x$ are

\begin{equation*}
\left\langle B_{x}\left( \vec{r}\right) B_{x}\left( \vec{r}\right)
\right\rangle =\frac{123\pi\hbar a^{2}}{256d^{5}}\Ima \left[ \frac{2}{ka} 
\frac{J_{1}\left( ka\right) }{J_{0}\left( ka\right) }\right] \coth\left( 
\frac{\hbar\omega}{2k_{B}T}\right) ,
\end{equation*}
valid for any value of $\delta/a$ and for $a\gg \delta$ we have 
\begin{equation*}
\left\langle B_{x}\left( \vec{r}\right) B_{x}\left( \vec{r}\right)
\right\rangle =\frac{123\pi\hbar a^{4}}{1024~d^{5}\delta^{2}}\coth\left( 
\frac{\hbar\omega}{2k_{B}T}\right) ,
\end{equation*}
while for $i=j=y$ 
\begin{equation*}
\left\langle B_{y}\left( \vec{r}\right) B_{y}\left( \vec{r}\right)
\right\rangle =\frac{3\pi\hbar a^{2}}{32d^{5}}\Ima \left[ \frac{2}{ka} 
\frac{J_{1}\left( ka\right) }{J_{0}\left( ka\right) }\right] \coth\left( 
\frac{\hbar\omega}{2k_{B}T}\right) ,
\end{equation*}
valid for any value of $\delta/a$ and for $a\gg \delta$ 
\begin{align*}
B_{y}^{\left( ind\right) } & =\frac{\pi a^{2}\alpha_{y}m}{d^{5}}\frac{3\pi }{ 
16} \\
\left\langle B_{y}\left( \vec{r}\right) B_{y}\left( \vec{r}\right)
\right\rangle & =\frac{3\pi\hbar a^{4}}{128d^{5}\delta^{2}}\coth\left( \frac{ 
\hbar\omega}{2k_{B}T}\right) .
\end{align*}

These considerations lead to very subtantial anisotropy in the correlation
functions and in the relaxation times. \ We have that for $d\gg a$ 
\begin{equation*}
\left\langle B_{x}\left( \vec{r}\right) B_{x}\left( \vec{r}\right)
\right\rangle :\left\langle B_{y}\left( \vec{r}\right) B_{y}\left( \vec{r} 
\right) \right\rangle :\left\langle B_{z}\left( \vec{r}^{\prime }\right)
B_{z}\left( \vec{r}\right) \right\rangle =82:16:9.
\end{equation*}

\bigskip

\subsection{Distant Object}
We now treat the magnetic noise of a metallic object whose maximum linear
dimension is short compared with the distance to the qubit: $d\gg L.$ We consider a fictitious point magnetic dipole $\vec{m}$ at $\vec{r}^{\prime} $ and a magnetically polarizable metallic object at the origin. \
The observation point is $\vec{r}$. \ Since $L$ is small, we may take the field $\vec{B}^{\prime}$ at the object due to the test dipole to be uniform over the object. If we assume that the electrode is spherical and its dielectric function is isotropic then the magnetic polarizability can be written as $\beta_{jn}\left( \omega\right) =\delta_{jn}\beta\left(\omega\right) $ and Eq. (\ref{eq:bgbb}) gives the physical correlation
function: 
\begin{align*}
\left\langle B_{i}\left( \vec{r}\right) B_{k}\left( \vec{r}^{\prime }\right)
\right\rangle & =\hbar\coth\left( \frac{\hbar\omega}{2k_{B}T}\right) \Ima \beta\frac{3x_{j}^{\prime}x_{k}^{\prime}-\delta_{jk}r^{\prime2}}{r^{\prime5} 
}\frac{3x_{i}x_{j}-\delta_{ij}r^{2}}{r^{5}}
\end{align*}
This manifestly satisfies the Onsager relation 
\begin{equation*}
G_{ik}\left( \omega;\vec{r},\vec{r}^{\prime}\right) =G_{ki}\left( \omega; 
\vec{r}^{\prime},\vec{r}\right) .
\end{equation*}
The local noise at $\vec{r} = \vec{r}^{\prime }$ is 
\begin{equation}
\left\langle B_{i}\left( \vec{r}\right) B_{k}\left( \vec{r}^{\prime}=\vec{r} 
\right) \right\rangle =\hbar\coth\left( \frac{\hbar\omega}{2k_{B} T}\right) 
\Ima \left( \beta\right) \frac{3x_{i}x_{k} +\delta_{ik}r^{2}}{r^{8}}.
\label{eq:bpolar}
\end{equation}
The $r^{-6}$ dependence is familiar from the van der Waals force, which has
a similar physical origin.\ 

Thus the problem reduces to a calculation of $\Ima \left[ \beta\left(
\omega\right) \right] ,$ the dissipative part of the polarizability of the
electrode. \ For a spherical electrode with radius $a$ and conductivity $ 
\sigma,$ we have that \cite{LLECM} 
\begin{equation*}
\Ima \beta=-\frac{3\delta^{2}a}{4}\left[ 1-\frac{a\sinh\left(
2a/\delta\right) +\sin\left( 2a/\delta\right) }{\cosh\left( 2a/\delta
\right) -\cos\left( 2a/\delta\right) }\right] ,
\end{equation*}
which reduces when $\delta\gg a$ to 
\begin{equation*}
\Ima \beta=\frac{a^{5}}{15\delta^{2}}
\end{equation*}
and when $\delta\ll a$ to 
\begin{equation*}
\Ima \beta=\frac{3a^{2}\delta}{4}.
\end{equation*}

Notice that the \textit{anisotropy} in lifetimes of a qubit in the presence
of a spherical electrode is independent of $\beta$. \ If the qubit is
located at $\vec{r}=r\widehat{z},$ then

\begin{equation*}
\left\langle B_{x}\left( \vec{r}\right) B_{x}\left( \vec{r}^{\prime}=\vec{r} 
\right) \right\rangle =\left\langle B_{y}\left( \vec{r}\right) B_{y}\left( 
\vec{r}^{\prime}=\vec{r}\right) \right\rangle =\hbar\coth\left( \frac{ 
\hbar\omega}{2k_{B}T}\right) \frac{\Ima \left[ \beta\left( \omega\right)  
\right] }{r^{6}}
\end{equation*}

\begin{equation*}
\left\langle B_{z}\left( \vec{r}\right) B_{z}\left( \vec{r}^{\prime}=\vec{r} 
\right) \right\rangle =4\hbar\coth\left( \frac{\hbar\omega}{2k_{B} T}\right) 
\frac{\Ima \left[ \beta\left( \omega\right) \right] }{r^{6}}.
\end{equation*}
The anisotropy 
\begin{equation*}
\left\langle B_{z}\left( \vec{r}\right) B_{z}\left( \vec{r}\right)
\right\rangle =4\left\langle B_{x}\left( \vec{r}\right) B_{x}\left( \vec {r} 
\right) \right\rangle 
\end{equation*}
is stronger than in the half-space case. \ 

\section{Numerical Estimates}
\begin{figure}[h!]
	\centering
	\includegraphics[width=0.7\linewidth]{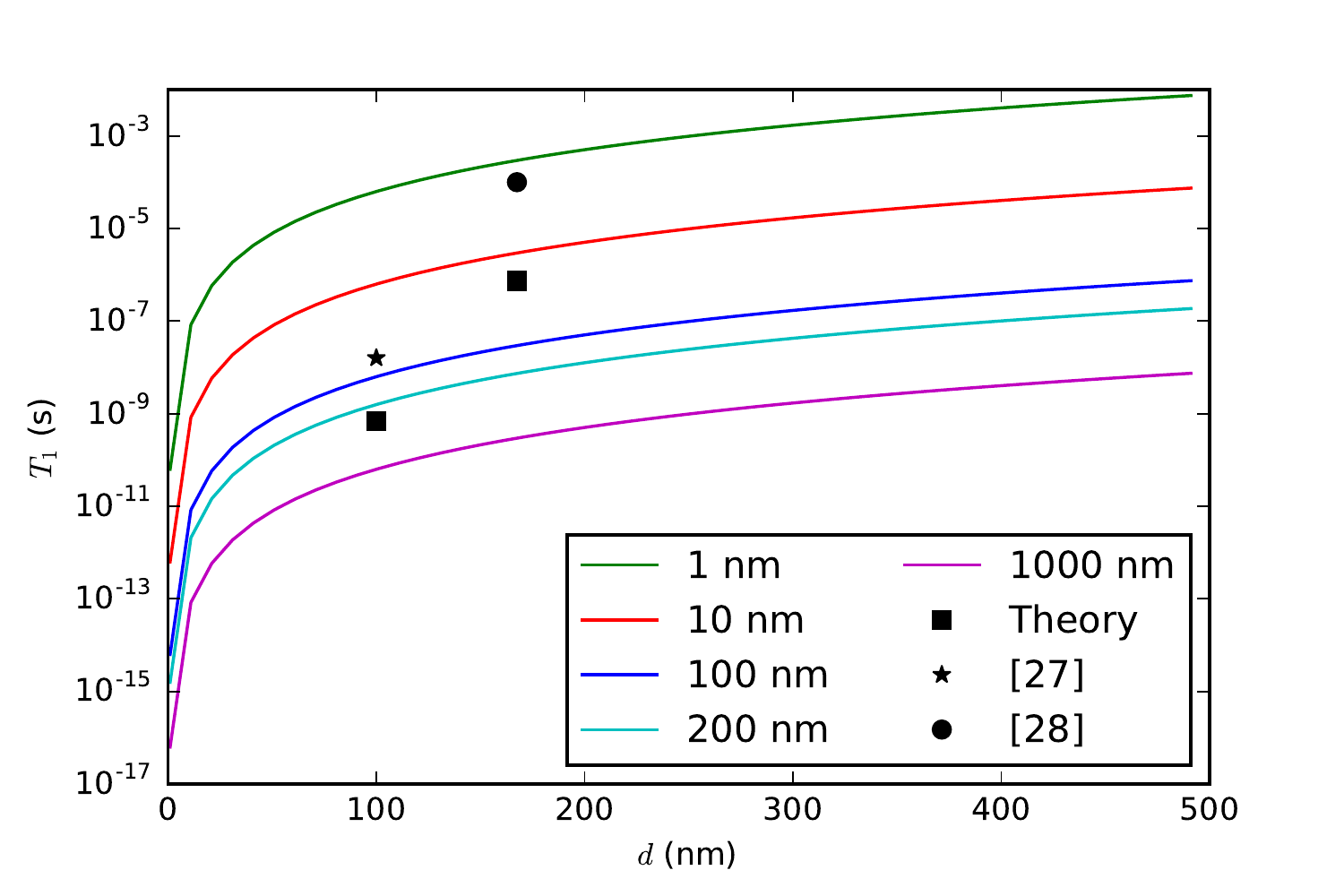}
	\caption{The charge qubit relaxation time $T_{1}$ as a function of the
		distance $d$ from the qubit to a planar metal gate for various values of the
		dot separation, as listed in the inset. \ The experimental data are taken
		from Ref. \cite{Petta,Gorman}. The theoretical predictions are
		indicated by solid squares. \ The qubit operating frequency is taken as $ 
		\protect\omega =10^{9}\;\text{s}^{-1}$, while the conductivities are roughly
		estimated as  $\protect\sigma =10^{16}\;\text{s}^{-1}.\text{ The temperature
			is }T=0.1~\text{K. }$}
	\label{fig:t1}
\end{figure}
In this section we provide some numerical estimates for the noise strength
and the resulting qubit relaxation times, which will allow us to evaluate
the relevance of EWJN for current experiments. \ We shall focus on the
half-space geometry, since this case is the important one for existing devices;
the greatest masses of metal in semicoductor qubit systems are usually in
global gates. \ 
\subsection{Charge Qubits}
The noise spectral energy density is of some interest. \ Taking $\omega
=10^{9}/s,$ $\sigma =10^{17}/s,$ we get $\delta =c/\sqrt{2\pi \sigma \omega } 
=12\times 10^{-4} ~ \text{cm}=12 \; \mu $ and we will only consider the regime $d\ll \delta .
$ \ The vacuum wavelength $\lambda =60$ cm is the longest length in the
problem and plays no role in our quasistatic regime. \ \ At a distance $d$
from a half space we find and $T=1$K and $\varepsilon _{d}=10$ we have$:$ \ 
\begin{equation*}
\left\langle E_{x}\left( d\right) E_{x}\left( d\right) \right\rangle
_{\omega }\approx \frac{k_{B}T}{8\pi \sigma d^{3}}=9\times 10^{-22}\frac{erg 
}{cm^{3}}s.
\end{equation*}

This noise will relax qubits. \ In Fig. \ref{fig:t1} we give numerical
estimates for $T_{1}$ of a charge qubit in a half-space geometry. \ The
curves are plotted using Eqs. (\ref{eq:T1exp}) and (\ref{eq:ezimage}) assuming a
point qubit. \  Each curve represents $T_{1}(d)$ for various values of the
distance $d$ from the half space and the dot separation $L$, the latter
being listed in the inset. \ We have assumed $\omega =10^{9}\;\text{s} 
^{-1},\sigma =10^{16}\;\text{s}^{-1},T=0.1\;\text{K}$. \ Indicated on the
figure are experimental values for $T_{1}$ and the predictions our model
makes based on estimates of the particular experiment's qubit and
surrounding geometry. The measured values are an order of magnitude or two
smaller than the predictions made by our model, indicating that EWJN is
probably not the dominant mechanism behind qubit relaxation in these
experiments. However, the estimates here are made with very limited
knowledge of the particular experimental values of $d$ and $L$, which are
normally not very accurately determined. \ Since $T_{1}\propto d^{3}/L^{2}$
a factor of $2$ could account for an order of magnitude correction. \ A
further serious source of uncertainty is that $\sigma $ is not measured and
generally is poorly known. \ If $\sigma $ is too large, the mean free path
is the electrons in the metal may become comparable to the gate dimensions,
invalidating the local electrodynamics used in this paper. \ \ These
considerations taken together mean that it is difficult to give a clear
evaluation of the role of EWJN in charge qubit experiments. \ In any case,
it seems safe to say that even rather minor improvements in other
decoherence mechanisms would make the EWJN mechanism competitive with the
others. 
\subsection{Spin qubits}
\begin{figure}[h!]
	\centering
	\includegraphics[width=0.7\linewidth]{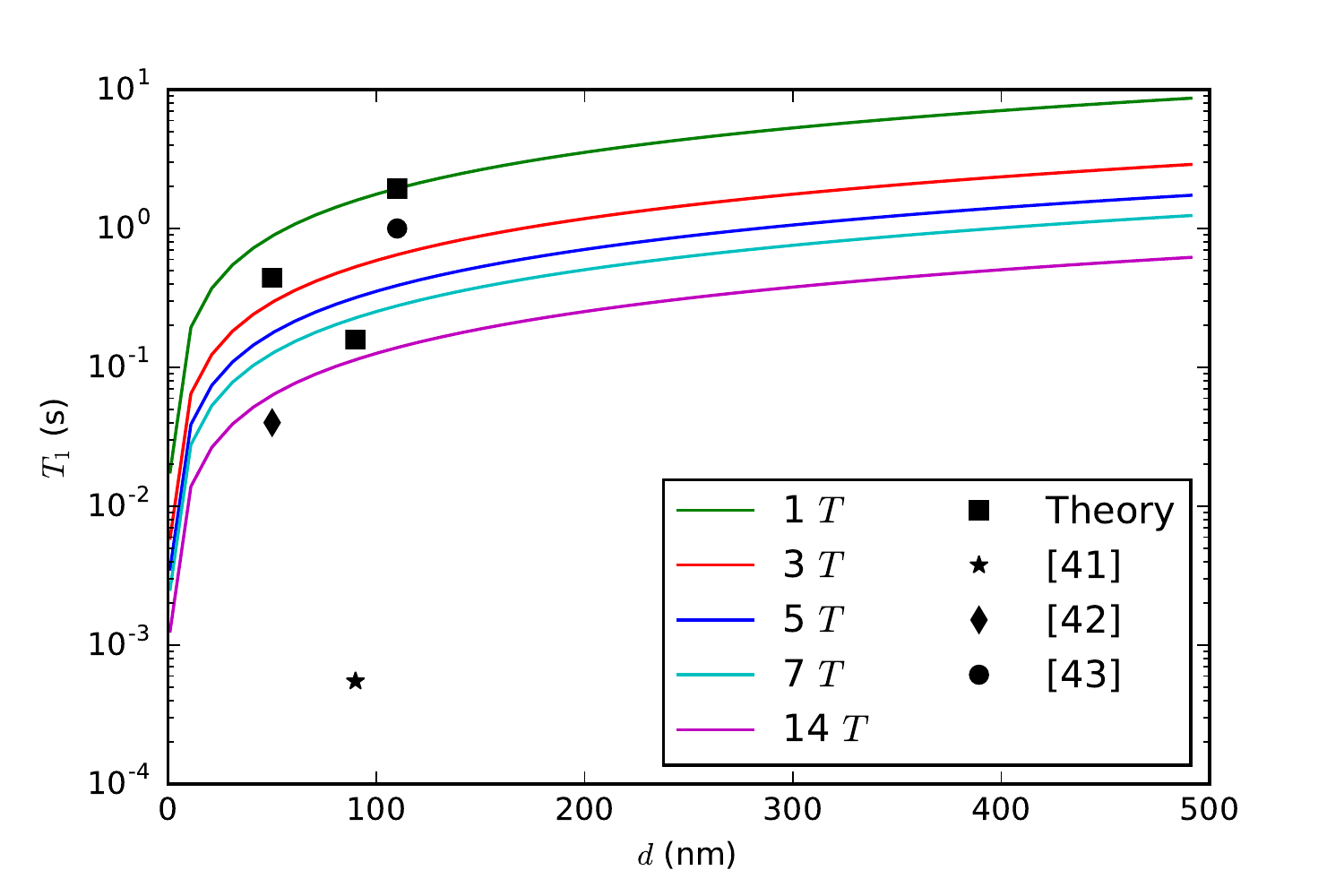}
	\caption{The spin qubit relaxation time $T_{1}$ as a function of the
		distance $d$ from the qubit to a planar metal gate for various values of the external magnetic field, as listed in the inset. \ The experimental data are taken
		from Ref. \cite{Xiao,Elzerman,Amasha}.  The theoretical predictions are
		indicated by solid squares. \ The conductivities are roughly
		estimated as  $\protect\sigma =10^{16}\;\text{s}^{-1}.\text{ The temperature
			is }T=0.1~\text{K. }$}
	\label{fig:t1mag}
\end{figure}

We can now repeat the numerical estimates for the noise strength and the
resulting qubit relaxation times for magnetic noise and spin qubits. 

The noise energy density is again of some interest. \ With $\omega =10^{9}/s,
$ $\sigma =10^{17}/s,$ $T=1$ K, $d=50$ nm from a half space we have$:$ 
\begin{equation*}
\left\langle B_{z}B_{z}\right\rangle _{\omega }\approx \frac{\pi
	k_{B}T\sigma }{dc^{2}}=\allowbreak 3.0\times 10^{-15}\frac{erg \dot s}{cm^{3}}
\end{equation*}
\

This noise will relax qubits. \ In Fig. \ref{fig:t1mag} we give numerical
estimates for $T_{1}$ of a spin qubit in a half-space geometry. \ The curves
are plotted using Eq. (\ref{eq:T1exp}) assuming a point
qubit. \  Each curve represents $T_{1}(d)$ for various values of the distance $d$ from the half space for a fixed $B$ field, which enters $T_1$ via $\hbar \omega = g \mu B$ with $g = 2$. \ We have assumed $\sigma =10^{16}\;\text{s}^{-1},T=0.1\;\text{K}$. \ Indicated on the
figure are experimental values for $T_{1}$ and the predictions our model
makes based on estimates of the particular experiment's qubit and
surrounding geometry. \ \ The geometries are somewhat better determined in
these experiments, meaning that that the main source of uncertainty is in the conductivity $\sigma ,$ which may differ from our assumption by an order of magnitude.

The experimental values shown in Fig. 7 are in small devices characterized by linear dimensions of order $100$ nm, but we note that in certain MOS devices the relevant distances can be closer to $10$ nm \cite{Angus,Hwang}. Other qubit architectures such as atom traps, ion traps, or superconductors, are generally considerably larger.  This makes it unlikely that EWJN plays a large role in the decoherence of these devices, since the power-law falloffs reduce the noise strength at the qubit positions.  This could change as these devices are miniaturized \cite{Follman, Cirone}.

\section{Conclusions}

Qubits with long relaxation times are necessary for quantum computation.
Most such devices are controlled electrically. This creates a control --
isolation dilemma: connections from the outside world are what make the
devices useful, but they are also sources of decoherence. In particular, one
may wish to place charge or spin qubits close to metallic device elements
used to confine or control the qubits. \ However, the fluctuating currents
and charges in metals give rise to noise that leaks out of the metal into
the surrounding region, decohereing the qubits. \ This is standard physics, (though not often treated in textbooks) and results for the noise spectral densities near a half plane are well known.  However, results for the more complicated geometries of real devices have not been available at all.  The results presented above represent a first step in the direction of repairing this situation.

Most importantly, we have given a streamlined method for the calculation of both
noise spectral densities and noise correlation functions. \ We have
presented new results for the spectral density of cylinders and distant
objects, and for the noise correlation functions for half spaces and distant
objects. \ The new method also enables us to give more qualitative, but
still useful, discussions of issues such as asperities on metal surfaces.

Numerical estimates of the effect of EWJN on qubits indicates that it is
proabably not a dominant effect on the current generation of charge qubit
devices. \ For spin qubits the situation is different. \ Experiments in
which the gates are close to the qubits may already be showing the effects
of EWJN. \  \  \  \  \  \  \ \ 
\section{Acknowledgments}
  We have benefited from discussions with Amrit Poudel, Luke Langsjoen, Andrea Morello, and John Nichol.  This work was supported by the ARO under award No. W911NF-12-0607.
\appendix
\section{Multipole moments in $T_1$ and $T_2$}

The expressions in Eqs. (\ref{eq:T1exp}) and (\ref{eq:t2exp}) can be
generalized to higher order multipole moments by keeping more terms in the
Taylor expansion of the electromagnetic potentials. Define the electric moments $q_i = e r_i$, $q_{ij} = e r_i r_j$ and magnetic dipole $m_i = \frac{e}{2mc} ((\vec{r} \times \vec{\Pi})_i + g S_i)$. We then have 
\begin{align*}
[H_q,p_i ] &= -\frac{i \hbar e }{m} \Pi_i \\ 
[H_q, q_{ij}] &= -\frac{i \hbar q }{m} \left( r_i \Pi_j + r_j \Pi_i - i \hbar
\delta_{ij} \right).
\end{align*}
We can find the quadrupole contribution by expanding Eq. (\ref{eq:hn}) and
keeping track of all first derivative terms giving us 
\begin{align*}
&\braket{0| H_n(t)|1} = - \frac{e}{mc} \left[ A_i(0,t)\braket{0| \Pi_i |1}
\right. \\
&~~~+ \left.\braket{0|\left(\nabla_j A_i(r,t)\right)_{r=0} r_j \Pi_i|1}  
\right] -\frac{e g}{2mc} B_i \braket{0| S_i|1} \\
&= i \frac{\omega}{c} \braket{0|p_i|1} A_i(0,t) - \frac{e}{mc} \langle 0| 
\frac{1}{2} [ \nabla_j A_i]_{r=0} (r_i \Pi_j + r_j \Pi_i)|1\rangle \\
& ~~~- \frac{e}{2mc} (\nabla \times \vec{A})_{k, r=0} \langle
0|\epsilon_{ijk}r_j\Pi_i |1\rangle - \frac{e g}{2mc}B_i \braket{0| S_i|1} \\
&= i \frac{\omega}{c} \left(\braket{0|p_i|1} A_i(0,t) + \frac{1}{2} [
\nabla_j A_i]_{r=0} \braket{0|q_{ij}|1}\right) \\
& ~~~- B_k(0,t) \braket{0|m_k |1}.
\end{align*}
We have employed the vector identity
\begin{align*}
\left[\nabla_j A_i(r,t)\right]_{r=0} r_j \Pi_i &= \frac{1}{2} [ \nabla_j
A_i]_{r=0} (r_i \Pi_j + r_j \Pi_i) \\
& - \frac{1}{2} \epsilon_{ijk} (\nabla \times \vec{A})_{k, r=0} r_j \Pi_i.
\end{align*}
Now we can work out an expression for $T_1$ using Eq. (\ref{eqn:t1}) 
\begin{align*}
\frac{1}{T_1} = \frac{1}{\hbar^2} & \left[ \braket{p_i} \braket{p_l}^*  
\braket{E_i E_l}_{\omega} + \frac{1}{2} \braket{p_i} \braket{q_{lm}}^*  
\braket{E_i \nabla_m E_l}_{\omega} + \frac{1}{2} \braket{q_{ij}} \braket{p_l} 
^* \braket{\nabla_j E_i E_l}_{\omega} \right. \\
& + \braket{p_i} \braket{m_n}^* \braket{E_i B_n}_{\omega} + \braket{ m_k}  
\braket{p_l}^* \braket{ B_k E_l}_{\omega} + \braket{m_k} \braket{m_n}^*  
\braket{B_k B_n}_{\omega} \\
& - \left. \frac{1}{2} \braket{q_{ij}} \braket{m_n}^* \braket{ \nabla_j E_i
B_n}_\omega + \frac{1}{2} \braket{m_k}\braket{q_{lm}} ^* \braket{B_k
\nabla_m E_l }_\omega + \frac{1}{4} \braket{q_{ij}} \braket{q_{lm}}^*  
\braket{\nabla_j E_i \nabla_m E_l} \right] .
\end{align*}
A naive application of the analysis from the preceeding calculation would indicate that $E$-field noise will not contribute to diagonal elements of $H_n(t)$, but this is due to the incomplete application of the gauge condition $\phi = 0$. If we begin with the gauge-invariant Schr\"{o} 
dinger equation with an arbitrary scalar potential $\phi(r,t)$ and vector
potential obeying $\nabla \cdot \vec{A} = 0$ and eliminate the residual
gauge freedom via $\vec{A}^{\prime }(r,t) = \vec{A}(r,t) + \nabla
\alpha(r,t) $, $\phi^{\prime }(r,t) = \phi(r,t) - \dot{\alpha}(r,t)= 0$ and $ 
\psi^{\prime} = e^{ -i e \alpha/\hbar}\psi(r,t)$ we find that the wave function
obeys 
\begin{equation*}
i \hbar \dot{\psi} = \left( e^{i e \alpha/\hbar} H^{\prime -i e
	\alpha/\hbar} - e \dot{\alpha}\right) \psi.
\end{equation*}
The Hamiltonian $H^{\prime }$ in the gauge with no scalar potential is
complemented by the gauge fixing term that retains the electric field
contribution in the equations of motion. The operator we need in Eq. (\ref 
{eq:t2s}) can be expanded in Taylor series as 
\begin{align}
H_n(t) &= - \frac{e}{mc} \left( A_i(0,t) + \left[\nabla_j A_i(r,t)\right] 
_{r=0} r_j + \dots \right) \Pi_i  \notag \\
&- \frac{e g}{2mc} B_i S_i. - e \left( \dot{\alpha}(0,t) + \nabla_j\dot{ 
	\alpha}(0,t) r_j + \dots \right).
\end{align}
Now turning to the relevant matrix elements of Eq. (\ref{eq:hn}) with the
gauge term (equivalently, the scalar potential), we begin by treating the
vector potential terms 
\begin{align*}
-\frac{e}{mc}\braket{1| A_i(0,t) \Pi_i |1} &= \frac{i}{\hbar c} A_i(0,t)
\left( \braket{1| p_i H_q |1} - \braket{1| H_q p_i |1} \right) \\
&= \frac{i }{\hbar c} A_i (\epsilon_1 - \epsilon_1) \braket{1|p_i|1} = 0, \\
-\frac{e}{mc}\braket{1| \nabla_j A_i r_j \Pi_i|1} &= - \frac{e}{mc} \left[ 
\frac{1}{2} \nabla_j A_i \braket{1|r_i \Pi_j + r_j \Pi_i|1} \right. \\
& \hspace{1cm}\;+\left. \frac{1}{2} B_k \braket{1|l_k|1} \right] \\
&= - \frac{e}{mc} \left(\frac{i \hbar}{2} \nabla_j A_i \delta_{ij} \right) - 
\frac{e}{2mc} B_k \braket{1|l_k|1} \\
&= - \frac{e}{2mc} B_k \braket{1|l_k|1}.
\end{align*}
The last equality follows from $\nabla \cdot \vec{A} = 0$. We have $E_i(r,t) =
- \nabla_i\dot{\alpha}(r,t)$.
Using the same methods we obtain an expression for the integral kernel $ 
S(\omega)$ for $T_2$ to quadrupole order. 
\begin{align*}
S(\omega) &= \frac{1}{\hbar^2} \left[\braket{B_i(t)B_j(0)}_\omega \Delta m_i
\Delta m_j -\braket{B_i(t)E_j(0)}_\omega \Delta m_i \Delta p_j -  
\braket{E_i(t)B_j(0)}_\omega \Delta p_i \Delta m_j \right. \\
& + \braket{E_i(t)E_j(0)} 
_\omega \Delta p_i \Delta p_j + \frac{1}{2} \left(\braket{\nabla_i E_j(t) B_k(0)}_\omega\Delta q_{ij}
\Delta m_k + \braket{\nabla_i E_j(t) E_k(0)}_\omega \Delta q_{ij} \Delta p_k \right.\\
& \left. + \braket{B_i(t) \nabla_j E_k(0) } \Delta m_i\Delta q_{jk} + \braket{E_i(t)
\nabla_j E_k(0) }_\omega \Delta p_i\Delta q_{jk} \right) \left. \frac{1}{4} \braket{\nabla_i E_j(t) \nabla_k E_l(0)}_\omega \Delta
q_{ij} \Delta q_{kl} \right]
\end{align*}

\section{Spectral Density Tensors}

Here we include the details and off-diagonal components of the noise spectral density tensor for the simple geometries treated in the main body of the paper. 

\subsection{Electric Noise}
Place a fictitious electric dipole moment $\vec{p}$ at the point $\vec{r}^{\prime}=\left( 0,0,d\right)$ in the half-space geometry. \ The electric field in free space would be
\begin{equation}
E_{j}^{\left( ed\right) }\left( \vec{r}\right) =-\frac{\partial}{\partial
	x_{j}}\vec{p}\cdot\nabla\frac{1}{\left\vert \vec{r}-\vec{r}^{\prime
	}\right\vert }=-p_{k}\frac{\partial}{\partial x_{j}}\frac{\partial}{\partial
	x_{k}}\frac{1}{\left\vert \vec{r}-\vec{r}^{\prime}\right\vert },
\label{eq:ed}
\end{equation}
which satisfies $\nabla\cdot\vec{E}^{\left( ed\right) }=4\pi\rho$ with $ 
\rho=-\vec{p}\cdot\nabla\delta^{3}\left( \vec{r}-\vec{r}_{0}\right) .$

We will represent this fictitious field by using the identity 
\begin{equation*}
\frac{1}{\left\vert \vec{r}-\vec{r}^{\prime}\right\vert }=\frac{1}{2\pi}\int 
\frac{d^{2}q}{q}~e^{i\vec{q}\cdot\vec{\rho}}e^{-q\left\vert z-d\right\vert },
\end{equation*}
where $\vec{q}=\left( q_{x},q_{y}\right) $ and $\vec{\rho}=\left( x,y\right)
.$ \ Thus 
\begin{equation*}
E_{j}^{\left( ed\right) }\left( \vec{r}\right) =-\frac{p_{k}}{2\pi}\int\frac{ 
	d^{2}q}{q}\frac{\partial}{\partial x_{j}}\frac{\partial}{\partial x_{k}}~e^{i 
	\vec{q}\cdot\vec{\rho}}e^{-q\left\vert z-d\right\vert }.
\end{equation*}
The induced field for $z>0$ is expanded as

\bigskip 
\begin{equation*}
E_{j}^{\left( ind\right) }\left( \vec{r}\right) =-\frac{p}{2\pi}\int
d^{2}q~f_{j}\left( \vec{q}\right) e^{i\vec{q}\cdot\vec{\rho}}e^{-qz},
\end{equation*}
and the Maxwell equations imply 
\begin{equation*}
\nabla^{2}\vec{E}^{\left( ind\right) }=0,~\nabla\cdot\vec{E}^{\left(
	ind\right) }=0,~z>0,
\end{equation*}
so 
\begin{align*}
q & =\sqrt{q_{x}^{2}+q_{y}^{2}} \\
iq_{x}f_{x}+iq_{y}f_{y} & =qf_{z}.
\end{align*}
The induced field for $z<0$ is defined by

\bigskip 
\begin{equation*}
E_{j}^{\left( ind\right) }\left( \vec{r}\right) =-\frac{p}{2\pi}\int
d^{2}q~g_{j}\left( \vec{q}\right) e^{i\vec{q}\cdot\vec{\rho}}e^{\alpha z}
\end{equation*}
and we have 
\begin{equation*}
\nabla^{2}\vec{E}^{\left( ind\right) }+2i\delta^{-2}\vec{E}^{\left(
	ind\right) }=0,~\nabla\cdot\vec{E}^{\left( ind\right) }=0,~z<0
\end{equation*}
and $ \Rea \alpha>0$ and so 
\begin{align*}
\alpha^{2} & =q_{x}^{2}+q_{y}^{2}-2i\delta^{-2}=q^{2}-2i\delta^{-2} \\
iq_{x}g_{x}+iq_{y}g_{y} & =-\alpha g_{z}.
\end{align*}

The tangential component of $\vec{E}$ is continuous but the normal component
satisfies $\vec{E}_{norm,out}=\varepsilon\vec{E}_{norm,in}\approx\left( 4\pi
i\sigma/\omega\right) \vec{E}_{norm,in},$ so $\left\vert \vec{E} 
_{norm,out}\right\vert \gg \left\vert \vec{E}_{norm,in}\right\vert .$ \ $\vec { 
	B}$ is continuous.
The fictitious dipole $\vec{p} = p \hat{z}$ produces a field for $0<z<d$ 
\begin{align*}
E_{x}^{\left( ed\right) }\left( \vec{r}\right) & =-\frac{p}{2\pi}\int
d^{2}q~iq_{x}~e^{i\vec{q}\cdot\vec{\rho}}e^{q\left( z-d\right) } \\
E_{y}^{\left( ed\right) }\left( \vec{r}\right) & =-\frac{p}{2\pi}\int
d^{2}q~iq_{y}~e^{i\vec{q}\cdot\vec{\rho}}e^{q\left( z-d\right) } \\
E_{z}^{\left( ed\right) }\left( \vec{r}\right) & =-\frac{p}{2\pi}\int
d^{2}q~q~e^{i\vec{q}\cdot\vec{\rho}}e^{q\left( z-d\right) },
\end{align*}
and the induced field is defined by 
\begin{align*}
\vec{E}^{\left( ind\right) } & =-\frac{p}{2\pi}\int d^{2}q~\vec{f}\left( 
\vec{q}\right) e^{i\vec{q}\cdot\vec{\rho}-qz}~\text{for }z>0\text{ and} \\
\vec{E}^{\left( ind\right) } & =-\frac{p}{2\pi}\int d^{2}q~\vec{g}\left( 
\vec{q}\right) e^{i\vec{q}\cdot\vec{\rho}+\alpha z}~\text{for }z<0.
\end{align*}

The boundary conditions yield 
\begin{align*}
iq_{x}e^{-qd}+f_{x} & =g_{x} \\
iq_{y}e^{-qd}+f_{y} & =g_{y} \\
qe^{-qd}+f_{z} & =\left( \varepsilon_{m}/\varepsilon_{d}\right) g_{z} \\
iq_{x}f_{x}+iq_{y}f_{y} & =qf_{z} \\
iq_{x}g_{x}+iq_{y}g_{y} & =-\alpha g_{z}.
\end{align*}
The solution is 
\begin{equation*}
\left( f_{x},f_{y},f_{z}\right) =\left( -iq_{x},-iq_{y},q\right) e^{-qd} 
\frac{1-\left( \varepsilon_{m}/\varepsilon_{d}\right) q/\alpha }{1+\left(
	\varepsilon_{m}/\varepsilon_{d}\right) q/\alpha},
\end{equation*}
which gives us an integral expression for the induced field and thus \ref{eq:extEzz}.
For $\vec{p} = p \hat{x}$ the dipole produces a field for $0<z<d$ 
\begin{align*}
E_{x}^{\left( ed\right) }\left( \vec{r}\right) & =\frac{p}{2\pi}\int \frac{ 
	d^{2}q~q_{x}^{2}}{q}~e^{i\vec{q}\cdot\vec{\rho}}e^{q\left( z-d\right) } \\
E_{y}^{\left( ed\right) }\left( \vec{r}\right) & =\frac{p}{2\pi}\int \frac{ 
	d^{2}q~q_{x}q_{y}}{q}~e^{i\vec{q}\cdot\vec{\rho}}e^{q\left( z-d\right) } \\
E_{z}^{\left( ed\right) }\left( \vec{r}\right) & =-\frac{ip}{2\pi}\int\frac{ 
	d^{2}q~q_{x}q}{q}~e^{i\vec{q}\cdot\vec{\rho}}e^{q\left( z-d\right) }.
\end{align*}
For $d \ll \delta$ we find
\begin{align*}
\Ima \vec{E}^{\left( ind\right) }\left( \vec{r}\right) & =-\frac{p}{2\pi 
}\frac{\omega \varepsilon _{d}}{2\pi \sigma }\frac{\partial }{\partial x} 
\nabla \int d^{2}q~\frac{1}{q}e^{-q\left( d+z\right) }e^{iq_{x}x+iq_{y}y} \\
& =-\frac{p}{2\pi }\frac{\omega \varepsilon _{d}}{\sigma }\frac{\partial }{ 
	\partial x}\nabla \frac{1}{\left[ \left( z+d\right) ^{2}+\rho ^{2}\right]
	^{1/2}} \\
& =\frac{p}{2\pi }\frac{\omega \varepsilon _{d}}{\sigma }\nabla \frac{x}{ 
	\left[ \left( z+d\right) ^{2}+\rho ^{2}\right] ^{3/2}}.
\end{align*} 
So, for example, 
\begin{equation*}
\Ima E_{x}^{\left( ind\right) }\left( \vec{r}\right) =-\frac{p}{2\pi } 
\frac{\omega \varepsilon _{d}}{\sigma }\frac{2x^{2}-\left( z+d\right)
	^{2}-y^{2}}{\left[ \left( z+d\right) ^{2}+\rho ^{2}\right] ^{5/2}}.
\end{equation*}
In the regime $d \gg \delta$ we find 
\begin{align*}
\Ima \vec{E}^{\left( ind\right) }\left( \vec{r}\right) & =-\frac{p}{2\pi 
}\frac{\omega \varepsilon _{d}}{2\pi \sigma \delta }\frac{\partial }{ 
\partial x}\nabla \int d^{2}q~\frac{1}{q^{2}}e^{-q\left( d+z\right)
}e^{iq_{x}x+iq_{y}y} \\
& =\frac{p}{2\pi }\frac{\omega \varepsilon _{d}}{\sigma \delta }\nabla
\left\{ \frac{x}{\rho ^{2}}\left[ 1-\frac{d+z}{\left[ \left( z+d\right)
	^{2}+\rho ^{2}\right] ^{1/2}}\right] \right\} ,
\end{align*} 
which gives us the correlation functions presented in the main text. 

For $\vec{p} = p \hat{z}$ and $d \ll  \delta$ the off-diagonal components are: 
\begin{equation}
\left\langle E_{x}\left( \vec{r}=\left( \vec{\rho},z\right) \right)
E_{z}\left( \vec{r}^{\prime}=\left( 0,0,d\right) \right) \right\rangle
_{\omega}=\frac{3\hbar\omega\varepsilon_{d}}{2\pi\sigma}\frac{x\left(
d+z\right) }{\left[ \left( d+z\right) ^{2}+\rho^{2}\right] ^{5/2}}\coth\frac{ 
\hbar\omega}{2k_{B}T},  \label{eq:exz}
\end{equation}
\begin{equation*}
\left\langle E_{y}\left( \vec{r}=\left( \vec{\rho},z\right) \right)
E_{z}\left( \vec{r}^{\prime}=\left( 0,0,d\right) \right) \right\rangle
_{\omega}=\frac{3\hbar\omega\varepsilon_{d}}{2\pi\sigma}\frac{y\left(
d+z\right) }{\left[ \left( d+z\right) ^{2}+\rho^{2}\right] ^{5/2}}\coth\frac{ 
\hbar\omega}{2k_{B}T}.
\end{equation*}
When $d \gg  \delta$ we have 
\begin{equation}
\left\langle E_{x}\left( \vec{r}=\left( \vec{\rho},z\right) \right)
E_{z}\left( \vec{r}^{\prime}=\left( 0,0,d\right) \right) \right\rangle
_{\omega}=\frac{\hbar\omega\varepsilon_{d}}{2\pi\sigma\delta}\frac{x}{\left[
\left( d+z\right) ^{2}+\rho^{2}\right] ^{3/2}}\coth\frac{\hbar\omega }{ 
2k_{B}T},  \label{eq:ezxdg}
\end{equation}
\begin{equation*}
\left\langle E_{y}\left( \vec{r}=\left( \vec{\rho},z\right) \right)
E_{z}\left( \vec{r}^{\prime}=\left( 0,0,d\right) \right) \right\rangle
_{\omega}=\frac{\hbar\omega\varepsilon_{d}}{2\pi\sigma\delta}\frac{y}{\left[
\left( d+z\right) ^{2}+\rho^{2}\right] ^{3/2}}\coth\frac{\hbar\omega }{ 
2k_{B}T}.
\end{equation*}

Now we turn to the solution for $\vec{p}=p\hat{x}$ and $d\ll \delta $. The
off-diagonal components are 
\begin{equation}
\left\langle E_{y}\left( \vec{r}\right) E_{x}\left( \vec{r}=\vec{r}^{\prime
}\right) \right\rangle _{\omega }=\frac{3\hbar \omega \varepsilon _{d}}{2\pi
\sigma }\frac{xy}{\left[ \left( z+d\right) ^{2}+\rho ^{2}\right] ^{5/2}} 
\coth \frac{\hbar \omega }{2k_{B}T}
\end{equation} 
\begin{equation*}
\left\langle E_{z}\left( \vec{r}\right) E_{x}\left( \vec{r}=\vec{r}^{\prime
}\right) \right\rangle _{\omega }=-\frac{3\hbar \omega \varepsilon _{d}}{ 
2\pi \sigma }\frac{x\left( d+z\right) }{\left[ \left( z+d\right) ^{2}+\rho
^{2}\right] ^{5/2}}\coth \frac{\hbar \omega }{2k_{B}T},
\end{equation*} 
and comparison with Eq. (\ref{eq:exz}) shows that the Onsager relation is
satisfied.

The off-diagonal components when $d\gg \delta $ are: 
\begin{align}
\left\langle E_{y}\left( \vec{r}\right) E_{x}\left( \vec{r}^{\prime }\right)
\right\rangle _{\omega }& =-\frac{\hbar }{2\pi }\frac{\omega \varepsilon _{d} 
}{\sigma \delta }\coth \frac{\hbar \omega }{2k_{B}T}\times  \\
& \frac{xy}{\rho ^{2}}\left\{ \frac{2}{\rho ^{2}}\left[ 1-\frac{d+z}{\left[
\left( z+d\right) ^{2}+\rho ^{2}\right] ^{1/2}}\right] -\frac{d+z}{\left[
\left( z+d\right) ^{2}+\rho ^{2}\right] ^{3/2}}\right\} .  \notag
\end{align} 
\begin{equation*}
\left\langle E_{z}\left( \vec{r}\right) E_{x}\left( \vec{r}^{\prime }\right)
\right\rangle _{\omega }=-\frac{\hbar }{2\pi }\frac{\omega \varepsilon _{d}}{ 
\sigma \delta }\frac{x}{\left[ \left( z+d\right) ^{2}+\rho ^{2}\right] ^{3/2} 
}\coth \frac{\hbar \omega }{2k_{B}T},
\end{equation*} 
and comparison with Eq. (\ref{eq:ezxdg}) shows that the Onsager relation is satisfied.

For the distant object geometry Since $L$, the qubit size, is small, we may take the field $\vec{E}^{\prime}$ at the
electrode due to the test dipole to be uniform over the object. \ It is given by 
\begin{align*}
E_{j}^{\prime}\left( \vec{r}=0\right) & =p_{k}~\partial_{j}\partial _{k} 
\frac{1}{\left\vert \vec{r}^{\prime}\right\vert } \\
& =p_{k}\frac{3x_{j}^{\prime}x_{k}^{\prime}-\delta_{jk}r^{\prime2}}{ 
	r^{\prime5}} = p_k f_{jk}(\vec{r}),
\end{align*}
where we have defined the dipole function 
\begin{equation*}
f_{ij}\left( \vec{r}\right) \equiv\frac{3x_{i}x_{j}-\delta_{ij}r^{2}}{r^{5}}.
\end{equation*}
We shall take only the first term in the multipole expansion of the field
produced by the object. \ We will write this dipole as $\vec{p}^{\left(
	el\right) }$ (``el" for ``electrode".) \ \ It can be written as $p_{j}^{\left(
	el\right) }\left( \omega\right) =\alpha_{jn}\left( \omega\right)
E_{n}^{\prime}\left( \omega\right) $. \ At the observation point $\vec{r}$
the (again fictitious) field is 
\begin{align*}
E_{i}\left( \vec{r}\right) & =p_{j}^{\left( el\right) }\partial
_{i}\partial_{j}\frac{1}{\left\vert \vec{r}\right\vert } \\
& =\alpha_{jn}E_{n}^{\prime}f_{ij}\left( \vec{r}\right) \\
& =\alpha_{jn}p^{\prime }_{m}f_{mn}\left( \vec{r}^{\prime}\right)
f_{ij}\left( \vec{r}\right) ,
\end{align*}

This leads directly to 
\begin{equation}
\left\langle E_{i}\left( \vec{r}\right) E_{k}\left( \vec{r}^{\prime }\right)
\right\rangle =\hbar\coth\left( \frac{\hbar\omega}{2k_{B}T}\right) \Ima  
\left( \alpha_{jn}\right) f_{kn}\left( \vec{r}^{\prime }\right) f_{ij}\left( 
\vec{r}\right) .  \label{eq:eiekanis}
\end{equation}
Hence only the polarizability of the object is relevant in the problem. If we assume that the electrode is spherical and its dielectric function is
isotropic then $p_{j}^{\left( sph\right) }\left( \omega\right) =\alpha\left(
\omega\right) \delta_{jn} E_{n}^{\prime}\left( \omega\right) $ and 
\begin{align*}
E_{i}^{\left( sp\right) }\left( \vec{r}\right) & =\alpha\left( \omega\right)
p^{\prime }_k f_{jk}(\vec{r}^{\prime }) f_{ij}(\vec{r}) \\
& =\alpha\left( \omega\right) p^{\prime }_k \frac{9x_{i}x_{k}^{\prime}~\vec{r 
	}\cdot\vec {r}^{\prime}+\delta_{ik}r^{2}r^{\prime2}-3x_{i}x_{k}r^{ 
	\prime2}-3x_{i} ^{\prime}x_{k}^{\prime}r^{2}}{r^{5}r^{\prime5}}.
\end{align*}
Using Eq. (\ref{eq:de}), we have

\begin{equation*}
-\frac{\omega^{2}}{\hbar c^{2}}G_{ik}\left( \omega;\vec{r},\vec{r}^{\prime
}\right) =\alpha\left( \omega\right) f_{ij}\left( \vec{r}\right)
f_{jk}\left( \vec{r}^{\prime}\right) .
\end{equation*}
This manifestly satisfies the Onsager relation 
\begin{equation*}
G_{ik}\left( \omega;\vec{r},\vec{r}^{\prime}\right) =G_{ki}\left( \omega; 
\vec{r}^{\prime},\vec{r}\right) .
\end{equation*}
And we find
\begin{equation*}
\left\langle E_{i}\left( \vec{r}\right) E_{k}\left( \vec{r}^{\prime }\right)
\right\rangle=\hbar\coth\left( \frac{\hbar\omega}{2k_{B}T}\right) \Ima \left[ \alpha\left( \omega\right) \right] f_{kj}\left( \vec{r} 
^{\prime}\right) f_{ij}\left( \vec{r}\right).
\end{equation*}
\subsection{Magnetic Noise}
To find the noise tensor in the half space we place a magnetic dipole moment $\vec{m} = m \hat{z}$ at $\vec{r} = (d,0,0)$ in analogy to the electric field noise calculation. The magnetic field due to this fictitious dipole in free space would be
\begin{equation*}
B_{j}^{\left( md\right) }\left( \vec{r}\right) =m_{i}\frac{\partial }{ 
	\partial x_{i}}\frac{\partial}{\partial x_{j}}\frac{1}{\left\vert \vec {r}- 
	\vec{r}^{\prime}\right\vert },
	\end{equation*}
which satisfies $\nabla\times\vec{B}^{\left( md\right)}=4\pi\vec{J}/c$ and $\vec{J}=\vec{m}\times\nabla\delta^{3}\left(\vec{r}-\vec{r}^{\prime}\right). $ Proceeding analogously to Eq. (\ref{eq:ed}), we have
	\begin{equation*}
	B_{j}^{\left( md\right) }\left( \vec{r}\right) =\frac{m_{k}}{2\pi}\int\frac{ 
		d^{2}q}{q}\frac{\partial}{\partial x_{j}}\frac{\partial}{\partial x_{k}}~e^{i 
		\vec{q}\cdot\vec{\rho}}e^{-q\left\vert z-d\right\vert }.
		\end{equation*}
		where $\vec{q}=\left( q_{x},q_{y}\right) $ and $\vec{\rho}=\left( x,y\right) 
		$, and
		
		\begin{equation*}
		\vec{B}^{\left( ind\right) }\left( \vec{r}\right) =-\frac{m}{2\pi}\int
		d^{2}q~\left( -iq_{x},-iq_{y},q\right) e^{-qd}\frac{1-q/\alpha}{1+q/\alpha } 
		e^{iq_{x}x+iq_{y}y-qz}~\text{\ for }z>0.
		\end{equation*}	
For $d\ll \delta$ we find 
\begin{align*}
	\frac{1-q/\alpha}{1+q/\alpha}
	&=\frac{\sqrt{q^{2}-2i\delta^{-2}}-q}{\sqrt{q^{2}-2i\delta^{-2}}+q} \\
	&\approx\left(-\frac{i}{2q^{2}\delta^{2}}\right)
\end{align*}
and
				\begin{equation*}
				\vec{B}^{\left( ind\right) }\left( \vec{r}\right) =\frac{im}{4\pi \delta^{2}} 
				\int d^{2}q\frac{1}{q^{2}}~\left( -iq_{x},-iq_{y},q\right) e^{-q\left(
					z+d\right) }e^{iq_{x}x+iq_{y}y}~\text{\ for }z>0
					\end{equation*}
For $\vec{m} = m \hat{z}$ and $d \ll  \delta$ the off-diagonal components of
the noise tensor in the half space are: 
\begin{equation}
\left\langle B_{x}\left( \vec{r}\right) B_{z}\left( \vec{r}^{\prime }\right)
\right\rangle _{\omega}=\frac{\hbar}{2\delta^{2}}\frac{x}{\rho^{2}}\left\{ 1- 
\frac{\left( d+z\right) }{\left[ \left( d+z\right) ^{2}+\rho^{2}\right]
^{1/2}}\right\} \coth\frac{\hbar\omega}{2k_{B}T}
\end{equation}
\begin{equation*}
\left\langle B_{y}\left( \vec{r}\right) B_{z}\left( \vec{r}^{\prime }\right)
\right\rangle _{\omega}=\frac{\hbar}{2\delta^{2}}\frac{y}{\rho^{2}}\left\{ 1- 
\frac{\left( d+z\right) }{\left[ \left( d+z\right) ^{2}+\rho^{2}\right]
^{1/2}}\right\} \coth\frac{\hbar\omega}{2k_{B}T}.
\end{equation*}
In the regime where $d \gg  \delta$ we have 
\begin{equation}
\left\langle B_{x}\left( \vec{r}\right) B_{z}\left( \vec{r}^{\prime }\right)
\right\rangle _{\omega}=-\hbar\delta\coth\frac{\hbar\omega}{2k_{B}T}\times 
\left[ \frac{-12x\left( z+d\right) ^{2}+3x\rho^{2}}{\left[ \left( d+z\right)
^{2}+\rho^{2}\right] ^{7/2}}\right]
\end{equation}
\begin{equation*}
\left\langle B_{y}\left( \vec{r}\right) B_{z}\left( \vec{r}^{\prime }\right)
\right\rangle _{\omega}=-\hbar\delta\coth\frac{\hbar\omega}{2k_{B}T}\times 
\left[ \frac{-12y\left( z+d\right) ^{2}+3y\rho^{2}}{\left[ \left( d+z\right)
^{2}+\rho^{2}\right] ^{7/2}}\right] .
\end{equation*}
Now we turn to $\vec{m} = m \hat{x}$ and $d \ll  \delta$ where the
off-diagonal components are 
\begin{align}
\left\langle B_{y}\left( \vec{r}\right) B_{x}\left( \vec{r}^{\prime }\right)
\right\rangle _{\omega} & =\frac{\hbar}{2\delta^{2}}\coth \frac{\hbar\omega}{ 
2k_{B}T}\frac{\partial}{\partial x}\left[ \frac{y}{\rho }\int_{0}^{\infty}dq~ 
\frac{1}{q}e^{-q\left( z+d\right) }J_{1}\left( q\rho\right) \right] \\
& =\frac{\hbar}{2\pi}\coth\frac{\hbar\omega}{2k_{B}T}\frac{\partial}{ 
\partial x}\left[ \frac{y}{\rho^{2}}\left\{ \left[ \left( d+z\right)
^{2}+\rho ^{2}\right] ^{1/2}-\left( d+z\right) \right\} \right]  \notag \\
& =-\frac{\hbar}{4\pi\delta^{2}}\coth\frac{\hbar\omega}{2k_{B}T}\left[ \frac{ 
-2xy}{\rho^{4}}\left\{ \left[ \left( d+z\right) ^{2}+\rho ^{2}\right]
^{1/2}-\left( d+z\right) \right\} +\frac{xy}{\rho^{2}}\left[ \left(
d+z\right) ^{2}+\rho^{2}\right] ^{-1/2}\right]  \notag
\end{align}
\begin{align}
\left\langle B_{z}\left( \vec{r}\right) B_{x}\left( \vec{r}^{\prime }\right)
\right\rangle _{\omega} & =-\frac{\hbar}{2\delta^{2}}\coth \frac{\hbar\omega 
}{2k_{B}T}\frac{\partial}{\partial x}\frac{\partial}{\partial z} 
\int_{0}^{\infty}dq~\frac{1}{q^{2}}e^{-q\left( z+d\right) }J_{0}\left(
q\rho\right) \\
& =\frac{\hbar}{2\delta^{2}}\coth\frac{\hbar\omega}{2k_{B}T}\frac{\partial }{ 
\partial x}\int_{0}^{\infty}dq~\frac{1}{q}e^{-q\left( z+d\right)
}J_{0}\left( q\rho\right)  \notag \\
& =-\frac{\hbar}{2\delta^{2}}\coth\frac{\hbar\omega}{2k_{B}T}\frac{x}{\rho
^{2}}\left\{ \left[ \left( d+z\right) ^{2}+\rho^{2}\right] ^{1/2}-\left(
d+z\right) \right\} .  \notag
\end{align}

On the other hand when $d\gg \delta $ we find 
\begin{align}
\left\langle B_{y}\left( \vec{r}\right) B_{x}\left( \vec{r}^{\prime }\right)
\right\rangle _{\omega }& =-15\hbar \delta \coth \frac{\hbar \omega }{2k_{B}T 
}\frac{xy\left( d+z\right) }{\left[ \left( d+z\right) ^{2}+\rho ^{2}\right]
^{7/2}} \\
\left\langle B_{z}\left( \vec{r}\right) B_{x}\left( \vec{r}^{\prime }\right)
\right\rangle _{\omega }& =\hbar \delta \coth \frac{\hbar \omega }{2k_{B}T} 
\frac{-12x\left( d+z\right) ^{2}+3x\rho ^{2}}{\left[ \left( d+z\right)
^{2}+\rho ^{2}\right] ^{7/2}}.
\end{align}
The distant object geometry results can be obtained by placing a fictitious dipole near a magnetically polarizable electrode. Since $d \gg L$ we assume the field generated by this electrode is uniform over the qubit and  given by 
\begin{equation*}
B_{k}^{\prime}\left( 0\right) =m_{j}~f_{kj}\left( \vec{r}^{\prime}\right) ,
\end{equation*}
where again $f_{jk}\left( \vec{r}\right) =\left(
3r_{i}r_{j}-r^{2}\delta_{ij}\right) /r^{5}.$ \ We shall take only the first
term in the multipole expansion of the field produced by the object, which
is completely characterized by its dipole moment $\vec{m}^{\prime}$. \
Assuming linear response yields $m_{i}^{\prime}=\beta_{ij}B_{j}^{\prime}$,
where $\beta_{ij}$ is the magnetic polarizability of the object. \ At the
observation point $\vec{r}$ the (again fictitious) field is 
\begin{align*}
B_{i}\left( \vec{r}\right) & =m_{m}^{\prime}f_{im}\left( \vec{r}\right) \\
& =\beta_{mk}m_{j}~f_{kj}\left( \vec{r}^{\prime }\right) f_{im}\left( \vec{r} 
\right) ,
\end{align*}
and the prescription following Eq. \ref{eq:db} then gives the physical noise
function as 
\begin{equation*}
\left\langle B_{i}\left( \vec{r}\right) B_{j}\left( \vec{r}^{\prime }\right)
\right\rangle =\hbar\Ima \left( \beta_{mk}\right) ~f_{kj}\left( \vec{r} 
^{\prime}\right) f_{im}\left( \vec{r}\right) \coth\left(
\hbar\omega/k_{B}T\right) .
\end{equation*}
This leads directly to 
\begin{equation}
\left\langle B_{i}\left( \vec{r}\right) B_{k}\left( \vec{r}^{\prime }\right)
\right\rangle =\hbar\coth\left( \frac{\hbar\omega}{2k_{B}T}\right) \Ima  
\left( \beta_{jn}\right) f_{kn}\left( \vec{r}^{\prime }\right) f_{ij}\left( 
\vec{r}\right) .  \label{eq:bibkanis}
\end{equation}.

\end{document}